\documentclass[article]{emulateapj}
\usepackage{natbib}
\newcommand{\hi}{\mbox{\rm \ion{H}{1}}} 
\newcommand{\hii}{\mbox{\rm \ion{H}{2}}}
\newcommand{\htwo}{\mbox{\rm H$_2$}} 
\newcommand{\htwofir}{\mbox{\rm H$_2^{\rm FIR}$}}

\newcommand{\xcounits}{\mbox{cm$^{-2}$ (K km s$^{-1}$)$^{-1}$}}
\newcommand{\xcounitsfrac}{\ensuremath{\frac{\mbox{cm$^{-2}$}}{\mbox{K km
        s$^{-1}$}}}} 

\newcommand{\xco}{\mbox{$X_{\rm CO}$}}
\newcommand{\xgal}{\mbox{$X_{\rm Gal}$}}

\shorttitle{Structure Of A Low Metallicity GMC} \shortauthors{Leroy et al.}
\slugcomment{Accepted for publication in \emph{The Astrophysical Journal}}

\begin{document}
\title{The Structure of a Low-Metallicity Giant Molecular Cloud Complex}

\author{ Adam K. Leroy\altaffilmark{1}, Alberto Bolatto\altaffilmark{2},
  Caroline Bot \altaffilmark{3}, Charles W. Engelbracht\altaffilmark{4}, Karl
  Gordon\altaffilmark{5}, Frank P. Israel\altaffilmark{6}, M\'{o}nica
  Rubio\altaffilmark{7}, Karin Sandstrom\altaffilmark{8}, and Sne\v{z}ana
  Stanimirovi{\'c}\altaffilmark{9}}

\altaffiltext{1}{Max-Planck-Institut f{\" u}r Astronomie, D-69117 Heidelberg,
  Germany} 

\altaffiltext{2}{Department of Astronomy, University of Maryland, College
  Park, MD 20742} 

\altaffiltext{3}{UMR 7550, Observatoire Astronomiques de
  Strasbourg, Universite Louis Pasteur, F-67000 Strasbourg, France}

\altaffiltext{4}{Steward Observatory, University of Arizona, Tucson, AZ 85721}

\altaffiltext{5}{Space Telescope Science Institute, 3700 San Martin Drive, Baltimore, MD 21218, USA}

\altaffiltext{6}{Sterrewacht Leiden, Leiden University, PO Box 9513, 2300 RA
  Leiden}

\altaffiltext{7}{Departamento de Astronom\'{i}a, Universidad de Chile, Casilla
  36-D}

\altaffiltext{8}{Department of Astronomy and Radio Astronomy
  Laboratory, University of California, Berkeley, CA 94720}

\altaffiltext{9}{Astronomy Department, University of Wisconsin, Madison, 475
  N.  Charter St., WI 53711, USA}

\begin{abstract}
  To understand the impact of low metallicities on giant molecular cloud (GMC)
  structure, we compare far infrared dust emission, CO emission, and dynamics
  in the star-forming complex N83 in the Wing of the Small Magellanic
  Cloud. Dust emission (measured by {\em Spitzer} as part of the S$^3$MC and
  SAGE-SMC surveys) probes the total gas column independent of molecular line
  emission and traces shielding from photodissociating radiation. We calibrate
  a method to estimate the dust column using only the high-resolution {\em
    Spitzer} data and verify that dust traces the ISM in the \hi -dominated
  region around N83. This allows us to resolve the relative structures of
  \htwo , dust, and CO within a giant molecular cloud complex, one of the
  first times such a measurement has been made in a low-metallicity
  galaxy. Our results support the hypothesis that CO is photodissociated while
  H$_2$ self-shields in the outer parts of low-metallicity GMCs, so that
  dust/self shielding is the primary factor determining the distribution of CO
  emission. Four pieces of evidence support this view. First, the
  CO-to-\htwo\ conversion factor averaged over the whole cloud is very high
  $4$--$11 \times 10^{21}$~\xcounits , or $20$--$55$ times the Galactic
  value. Second, the CO-to-\htwo\ conversion factor varies across the complex,
  with its lowest (most nearly Galactic) values near the CO peaks. Third,
  bright CO emission is largely confined to regions of relatively high
  line-of-sight extinction, $A_V \gtrsim 2$~mag, in agreement with PDR models
  and Galactic observations. Fourth, a simple model in which CO emerges from a
  smaller sphere nested inside a larger cloud can roughly relate the
  \htwo\ masses measured from CO kinematics and dust.
\end{abstract}


\section{Introduction}
\label{INTRO}

Most star formation takes place in giant molecular clouds (GMCs). A
quantitative understanding of how local conditions affect the
structure and evolution of these clouds is key to link conditions in
the interstellar medium (ISM) to stellar output. Achieving such an
understanding is unfortunately complicated by the fact that H$_2$ does
not readily emit under the conditions inside a typical
GMC. Astronomers therefore rely on indirect tracers of H$_2$, most
commonly CO line emission and dust absorption or emission. These
tracers are also affected by environment, so that assessing the impact
of local conditions on GMC structure requires disentangling the effect
of these conditions on the adopted tracer from their effect on the
underlying distribution of H$_2$.

One way around this problem is to use several independent methods to measure
the structure of GMCs in extreme environments, inferring the state of H$_2$ by
comparing the results. Here we apply this approach to an active star-forming
region in the Small Magellanic Cloud (SMC).  Using far infrared (FIR) emission
measured by the {\em Spitzer} Survey of the SMC \citep[S$^3$MC][]{BOLATTO07}
and SAGE-SMC (``Surveying the Agents of a Galaxy's Evolution in the SMC'',
Gordon et al. in prep.), we derive the distribution of dust in the region. We
compare this to maps of CO and \hi\ line emission
\citep{BOLATTO03,STANIMIROVIC99}. Dust traces the total gas distribution ---
of which the atomic component is already known --- and offers a probe of
shielding from dissociating UV radiation. CO is the most common molecule after
H$_2$ (and the most commonly used tracer of molecular gas); understanding its
relation to H$_2$ in extreme environments is a long-standing goal. The CO line
also carries kinematic information that allows dynamical estimates of cloud
masses.

The SMC is of particular interest because the ISM in dwarf irregular galaxies
like the SMC contrast sharply with that of the Milky Way. They have low
metallicities \citep[e.g.,][]{LEE06}, correspondingly low dust-to-gas ratios
\citep[e.g.,][]{ISSA90,WALTER07}, and intense radiation fields
\citep[e.g.,][]{MADDEN06}. These factors should affect the formation and
structure of GMCs
\citep[e.g.,][]{MALONEY88,ELMEGREEN89,MCKEE89,PAPADOPOULOS02,PELUPESSY06}.
Unfortunately, it has proved extremely challenging to unambiguously observe
such effects because the inferred structure of GMCs depends sensitively on the
method used to trace \htwo .

Virial mass calculations reveal few differences between GMCs in dwarf galaxies
and those in the Milky Way. In this approach, one uses molecular line emission
to measure the size and line width of a GMC.  By assuming a density profile
and virial equilibrium, one can estimate the dynamical mass of the cloud
independent of its luminosity.  Recent studies find the ratio of virial mass
to luminosity for GMCs in other galaxies to be very similar to that observed
in the Milky Way \citep{WALTER01,WALTER02,ROSOLOWSKY03,
  BOLATTO03,ISRAEL03,LEROY06,BLITZ07,BOLATTO08}.  Further, the scaling
relations among GMC size, line width, and luminosity found in the Milky Way
\citep{LARSON81,SOLOMON87,HEYER08} seem to approximately apply to resolved CO
emission in other galaxies, even dwarf galaxies \citep{BOLATTO08}.

By contrast, observations of low metallicity galaxies that do not depend on
molecular line emission consistently suggest large reservoirs of
\htwo\ untraced by CO \citep[e.g.,][]{ISRAEL97, MADDEN97, PAK98, BOSELLI02,
  GALLIANO03,RUBIO04, LEROY07, BOT07}.  The most common manifestation of this
is an ``excess'' at FIR or sub-millimeter wavelengths with the following
sense: towards molecular peaks, there is more dust emission than one would
expect given the gas column estimated from \hi\ + CO.  \citet{ISRAEL97}
treated the abundance of \htwo\ as an unknown and used this excess to solve
for the CO-to-H$_2$ conversion factor. He found it to depend strongly on both
metallicity and radiation field.

These two sets of observations may be reconciled if CO is selectively
photodissociated in the outer parts of low-metallicity GMCs
\citep[e.g.][]{MALONEY88,ISRAEL88,BOLATTO99}, a scenario discussed
specifically for the SMC by \citet{ISRAEL86} and
\citet{RUBIO91,RUBIO93B}. This might be expected if \htwo\ readily
self-shields while CO is shielded from photodissociating radiation mostly by
dust, which is less abundant at low metallicities. In this case, CO emission
would trace only the inner parts of low-metallicity GMCs.

Observations of the Magellanic Clouds as part of the Swedish-ESO Submillimeter
Telescope (SEST) Key Programme \citep{ISRAEL93} support this idea: the surface
brightness of CO is very low in the SMC \citep{RUBIO91}; SMC clouds tend to be
smaller than their Milky Way counterparts, with little associated diffuse
emission \citep{RUBIO93B,ISRAEL03}; and the ratio of $^{13}$CO to $^{12}$CO
emission is lower in the Magellanic Clouds than in the Galaxy, suggesting that
clouds are more nearly optically thin \citep[][]{ISRAEL03}.

The SEST results are mainly indirect evidence. What is still needed is a
direct, {\em resolved} comparison between CO, dust, and H$_2$.  Because dust
emission offers a tracer of the total gas distribution that is independent of
molecular line emission
\citep[][]{THRONSON87,THRONSON88A,THRONSON88B,ISRAEL97}, it allows such a
test. If GMCs at low metallicity include envelopes of CO-free \htwo, then the
distribution of dust (after subtracting the dust associated with \hi ) should
be extended relative to CO emission.

\citet{LEROY07} attempted this measurement. They combined S$^3$MC with IRIS
data \citep{MIVILLE05} to derive the distribution of dust and compared this to
the NANTEN CO survey by \citet{MIZUNO01}. They derived a distribution of
\htwo\ $\sim 1.3$ times more extended than that of CO, suggesting that half of
the \htwo\ in the SMC may lie in envelopes surrounding the CO peaks. The
resolution of the CO and IRIS data limited this comparison to scales of
$\gtrsim 45$~pc. SMC GMCs are often much smaller than this
\citep[e.g.][]{RUBIO93B,MIZUNO01,ISRAEL03}. Therefore while this measurement
indicated that SMC GMC complexes may be immersed in a sea of CO-free cold gas,
it was not yet a true comparison of dust and CO on the scales of individual
GMCs.

Here, we focus on a single region, N83/N84 (hereafter simply N83).  This
isolated star-forming complex lies in the eastern Wing of the SMC and harbors
$\sim 10\%$ of that galaxy's total CO luminosity \citep{MIZUNO01}. Combining
FIR, CO, and \hi\ data we attempt to answer following questions:

\begin{enumerate}
\item What is the CO-to-H$_2$ conversion factor, \xco\ (i.e., the ratio of
  \htwo\ column density to CO intensity along a line of sight) in this region?
\item Is there evidence that CO is less abundant relative to H$_2$ (i.e., that
  \xco\ is higher or that there is H$_2$ without associated CO) in the outer
  parts of the cloud?
\item Is the distribution of CO consistent with dust shielding playing a key
  role in its survival?
\item Can dynamical masses measured from CO kinematics be brought into
  agreement with H$_2$ masses estimated from dust? What is the implied
  distribution of H$_2$?
\end{enumerate}

\noindent To meet these goals, we first estimate the dust optical depth at
160$\mu$m, $\tau_{160}$ (\S \ref{DUST}). We demonstrate that $\tau_{160}$
traces \hi\ column density in the (assumed) \hi -dominated ISM near N83, make
a self-consistent determination of the dust-to-gas ratio, and then combine
$\tau_{160}$ with the measured \hi\ column density to estimate the
\htwo\ column density in the star forming region (\S
\ref{GASANDDUST}). Finally, we combine the resulting maps of $\tau_{160}$ and
\htwo\ with CO and \hi\ data to answer the questions posed above (\S
\ref{STRUCTURE}).

\section{Data}
\label{DATA}

\begin{figure*}
\plotone{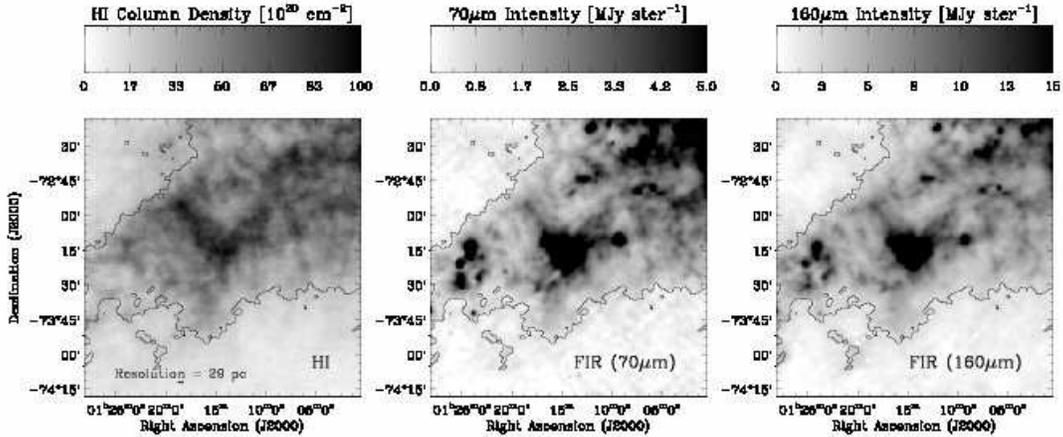}
\caption{\label{TWODEGREEFIELD}\hi\ (left) and FIR emission at 70 (middle) and
  160 $\mu$m (right) in a two degree wide field centered on N~83. A thin black
  contour outlines the region where we can clearly distinguish FIR emission
  from the background (see \S \ref{PROCESSING}).}
\end{figure*}

We use FIR imaging from two {\em Spitzer} surveys. S$^3$MC mapped 70 and
160$\mu$m emission from most active star forming regions in the SMC, including
N~83. More recently, SAGE-SMC observed a much larger area, including the
Magellanic Bridge and nearby emission-free regions. We use a combination of
these data sets carried out by Gordon et al. (in prep.) that dramatically
improves the quality of the 70$\mu$m image compared to S$^3$MC alone, thus
enabling this analysis. At $36\arcsec$ resolution, the noise ($1\sigma$) in
the {\em Spitzer} maps is $\sigma_{70} = 0.13$~MJy ster$^{-1}$ (70$\mu$m) and
$\sigma_{160} = 0.6$ MJy ster$^{-1}$ (160$\mu$m) in the neighborhood of N83.

We compare the {\em Spitzer} data to the IRIS 100$\mu$m image.  IRIS is a
re-processing of the IRAS data carried out by \citet{MIVILLE05}. These data
have $\sim 4.3\arcmin$ resolution.

\citet{BOLATTO03} used SEST to map CO $J=2\rightarrow1$ and
$J=1\rightarrow0$ emission from N83. The half-power beam width of SEST
was $23\arcsec$ ($J=2\rightarrow1$) and $45\arcsec$
($J=1\rightarrow0$). The maps that we use were convolved to lower
resolution during reduction and have final angular resolutions of
$38\arcsec$ ($J=2\rightarrow1$) and $55\arcsec$
($J=1\rightarrow0$). The noise in the velocity-integrated maps is
somewhat position-dependent. Over regions with significant emission
$1\sigma$ is typically $0.16$ K~km~s$^{-1}$ (CO $J=2\rightarrow1$) and
$0.22$~K~km~s$^{-1}$ (CO $J=1\rightarrow0$).

\citet{STANIMIROVIC99} imaged \hi\ 21-cm line emission across the whole SMC.
These data have angular resolution $98 \arcsec$ and sensitivity sufficient to
detect \hi\ emission along every line of sight within $\sim 1\arcdeg$ of N83.
We correct for \hi\ optical depth and self-absorption following \citet[][their
  Equation 6]{STANIMIROVIC99} based on the \hi\ absorption study by
\citet{DICKEY00}. The maximum correction factor near N83 is $\sim 1.3$.

To subtract emission associated with the Milky Way from the FIR maps (\S
\ref{PROCESSING}), we use the Parkes map of Milky Way \hi\ from
\citet{BRUNS05}. Galactic \hi\ is distinguished from SMC gas by its radial
velocity. These data have a resolution of $14\arcmin$.

We move all data to three astrometric grids: one covering the entire SMC, a
two degree wide field surrounding N~83 (Figure \ref{TWODEGREEFIELD}), and the
SEST field. In the SEST field, we use the kernels of \citet{GORDON08A} to
place the 70$\mu$m image at the 160$\mu$m resolution ($\sim 36\arcsec$), which
matches that of the SEST CO $J=2\rightarrow1$ data ($38\arcsec$) well. We also
convolve the 70 and 160$\mu$m maps to the $55\arcsec$ resolution of the SEST
CO $J=1\rightarrow0$ data. In the two degree field near N83, we use a Gaussian
kernel to place the 70 and 160$\mu$m data at the $98\arcsec$ resolution of the
\hi . Over the whole SMC, we degrade the 70 and 160$\mu$m images to the
$4.3\arcmin$ IRIS resolution.

\subsection{Additional Processing of the FIR Maps}
\label{PROCESSING}

For consistency among the 70, 100, and 160 $\mu$m data, we move flux densities
at 70 and 160 $\mu$m from the MIPS scale (which assumes $F_{\nu} \propto
\nu^2$ across the bandpass) to the IRAS scale (which assumes $F_{\rm \nu}
\propto \nu^{-1}$). We do so by dividing the $70$ map by $0.918$ and the
$160\mu$m map by $0.959$.

We subtract Milky Way foreground emission from the 100 and 160$\mu$m maps. We
estimate this from Galactic \hi\ assuming the average cirrus dust properties
measured by \citep{BOULANGER96}. At 100$\mu$m we use their fit directly; at
160$\mu$m we interpolate their fits assuming a typical cirrus dust temperature
($T=17.5$~K) and emissivity ($\beta =2$).

To refine the foreground subtraction, we assume that \hi\ and infrared
intensity from the SMC are correlated at a basic level. As the column density
of SMC \hi\ approaches 0, we expect the IR intensity of the SMC to also
approach 0. Therefore, we adjust the zero point of the IR maps using a fit of
IR intensity to $N\left(\hi \right)_{\rm SMC}$ where $N(\hi )_{\rm SMC} < 2
\times 10^{21}$~cm$^{-2}$ (we subtract the fitted $y$-intercept). This leads
us to add $0.3$~MJy~ster$^{-1}$ at 70~$\mu$m, subtract $4.4$~MJy~ster$^{-1}$
at 160$\mu$m, and subtract $0.5$~MJy~ster$^{-1}$ from the IRIS 100$\mu$m
map. These offsets are a natural consequence of the uncertainty in the
reduction and foreground subtraction (which must remove zodiacal light, Milky
Way cirrus, and any cosmic infrared background). Deviations from the average
cirrus properties are particularly common, being observed near a number of
galaxies by \citet{BOT09}.

Based on carrying out this exercise in several different ways, we estimate the
zero level of our maps to be uncertain by $0.25$~MJy~ster$^{-1}$ at 70$\mu$m
and $1$~MJy~ster$^{-1}$ at 160$\mu$m. We take these uncertainties into account
in our calculations (\S \ref{RANDOM_UNC}). To minimize their impact we only
consider lines of sight with intensities well above the background, by which
we mean $I_{70} > 0.5$~MJy~ster$^{-1}$ and $I_{160} > 2$~MJy~ster$^{-1}$ after
the foreground subtraction (i.e., twice the uncertainty in the background).

\subsection{A Word on Resolution}
\label{RESOLUTION}

In the rest of this paper we will combine the data described above in several
ways. Two of these combinations lead to maps combining data with different
resolutions. We comment on these here and the reader may wish to refer back to
this section while reading the paper.

First, we subtract a foreground component measured at $14\arcmin$ resolution
from IR maps with $4.3\arcmin$ and $\sim 36\arcsec$ (160$\mu$m)
resolution. Any small scale variation in the Milky Way cirrus will therefore
be left in our maps. This is only a concern in the diffuse region of the Wing
(and so only in \S \ref{HIVSFIR}). In N83 itself most lines of sight exhibit
FIR intensities $\gtrsim 10$ times higher than the foreground, so variations
in the foreground are not a concern.

Second, when estimating the distribution of \htwo\ in N83, we derive the total
amount of hydrogen ($\hi + \htwo $) along a line of sight and then subtract
the measured \hi\ column density. The total amount of hydrogen is based on FIR
dust emission, measured at $36\arcsec$ resolution (or $55\arcsec$ resolution
when we compare to the SEST CO $J=1\rightarrow 0$ map). The \hi\ column
density is measured at $98\arcsec$ resolution. We assume it to be smooth on
smaller scales, an assumption born out to some degree by the reasonable
correlation that we find between \htwo\ and CO. Nonetheless, the detailed
distribution of \htwo\ on scales less than $98\arcsec$ ($\sim 29$~pc) is
somewhat uncertain.

\section{Dust Treatment}
\label{DUST}

We use the optical depth at 160$\mu$m, $\tau_{160}$, as a proxy for
the amount of dust along a line of sight. For an optically thin
population of grains with an equilibrium temperature $T_{\rm dust}$,
$\tau_{160}$ is related to the measured $160\mu$m intensity,
$I_{160}$, by

\begin{equation}
\label{TAU160}
\tau_{160} = \frac{I_{160}}{B_{\nu} \left(T_{\rm dust}, 160\mu{\rm m}\right)}~.
\end{equation}

\noindent Here $B_{\nu} ~\left(T_{\rm dust}, \lambda \right)$ is the intensity
of a blackbody of temperature $T_{\rm dust}$ at wavelength $\lambda$.

Calculating $\tau_{160}$ thus requires estimating $T_{\rm dust}$. Because only
the 70 and 160$\mu$m maps have angular resolution appropriate to compare with
CO, we must do so using this combination.  Unfortunately, $I_{\rm 70}/I_{\rm
  160}$ does not trivially map to $T_{\rm dust}$ because the 70$\mu$m band
includes non-equilibrium emission from small grains
\citep[e.g.,][]{DESERT90,DRAINE07A,BERNARD08}. We therefore take an indirect
approach: we assume that most of the dust mass resides in large grains with
equilibrium temperature $T_{\rm dust}$ that contribute all of the emission at
100$\mu$m and 160$\mu$m. We use $I_{\rm 70}/I_{\rm 160}$ to estimate $I_{\rm
  100}/I_{\rm 160}$ and then solve for $T_{\rm dust}$ from

\begin{equation}
  \frac{I_{\rm 100}}{I_{\rm 160}} = \left(\frac{100}{160}\right)^{-1.5}~
  \frac{B_{\nu} \left(T_{\rm dust}, 100\mu{\rm m}\right)}{B_{\nu} \left(T_{\rm dust},160\mu{\rm m} \right)}~,
\end{equation}

\noindent which assumes that dust has a wavelength-dependent emissivity such
that $\tau_{\lambda} \propto \lambda^{-\beta}$ with $\beta = 1.5$.

We derive the relationship between $I_{\rm 70}/I_{\rm 160}$ and $I_{\rm
  100}/I_{\rm 160}$ at the $4.3\arcmin$ resolution of IRIS, where both colors
are known and exhibit a roughly 1-to-1 relation. We then assume this
relationship to apply to the smaller ($\sim 36\arcsec$) angular scales
measured only by the {\em Spitzer} data. Near N83, the two colors are related
by:

\begin{equation}
\label{POLYFIT}
\frac{I_{\rm 100}}{I_{\rm 160}} = 0.24 x^2 + 0.33 x + 0.45 ,~{\rm where}~x=\frac{I_{\rm 70}}{I_{\rm 160}}~.
\end{equation}

\noindent Note that this is not a general relation. It does not go through the
origin and is only 1-to-1 over a limited range of $I_{70}/I_{160} $; we fit
and apply over the range $I_{70}/I_{160} \sim 0.15$ -- $1.2$, where it is a
good description of the SMC.

\subsection{Motivation}
\label{DUST_MOTIVATION}

\begin{figure}
\plotone{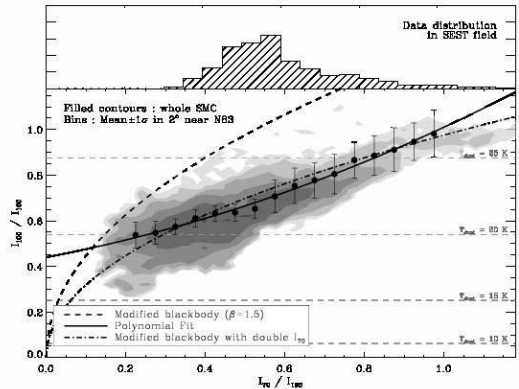}
\caption{\label{COLORCOLOR} FIR color-color plot for the SMC. The $x$-axis
  shows $I_{70}/I_{160}$, which is measured by {\em Spitzer} at high
  resolution but includes contamination by non-equilibrium emission. The
  $y$-axis shows $I_{100}/I_{160}$, which is only available at $4.3\arcmin$
  resolution but is more likely to trace exclusively equilibrium emission.
  Horizontal dashed lines show the temperatures associated with a few values
  of this color. Shaded contours show the distribution of data for the whole
  SMC; the lowest contour includes all data and the contour increment is a
  fact of $4$ in data density. Black circles show mean and $1\sigma$ scatter
  for data in a $2\arcdeg$ field centered on N83 (binned by
  $I_{70}/I_{160}$). The histogram above the plot shows the distribution of
  $I_{70}/I_{160}$ over the SEST field (i.e., N83 itself). The dashed curve
  shows the color-color relation expected for a modified blackbody, which is
  not a good description of the SMC. On the other hand, the solid and
  dash-dotted lines, which show the relations that we use in our analysis, can
  reasonably predict $I_{100}/I_{160}$ from $I_{70}/I_{160}$.}
\end{figure}

In assuming that $I_{\rm 100}/I_{\rm 160}$ traces $T_{\rm dust}$ or its more
sophisticated analogs \citep[e.g.,][]{DALE02,DRAINE07A}, we follow several
recent studies of the Magellanic Clouds
\citep[][]{BOT04,LEROY07,BERNARD08,GORDON08B}. \citet{SCHNEE05,SCHNEE06,SCHNEE08}
have demonstrated that a similar approach reproduces optical and near-IR
extinction in Galactic molecular clouds, though with some systematic
uncertainties.

Figure \ref{COLORCOLOR} motivates our use of $I_{\rm 70}/I_{\rm 160}$
($x$-axis) to predict $I_{\rm 100}/I_{\rm 160}$ ($y$-axis). Gray
contours show the distribution of data for the whole SMC. Bins (filled
circles) show data from a $2\arcdeg$ square field centered on N83
(i.e., Figure \ref{TWODEGREEFIELD}). Both near N83 and over the whole
SMC, the two colors show a reasonable correlation (rank correlation
coefficient $0.7$).

Figure \ref{COLORCOLOR} also motivates our {\em ad hoc} treatment of the
conversion between $I_{\rm 70}/I_{\rm 160}$ and $I_{\rm 100}/I_{\rm 160}$. A
single modified blackbody (the dashed line shows one with $\beta = 1.5$)
cannot simultaneously describe the SMC at 70, 100, and 160$\mu$m. The simplest
explanation is that $I_{\rm 100}/I_{\rm 160}$ traces $T_{\rm dust}$, while the
70$\mu$m band includes substantial non-equilibrium emission. We tested the
possibility of using the models of \citet{DRAINE07A}, which include the
effects of stochastic heating, to directly derive dust masses from
$I_{70}/I_{160}$. However, the currently available ``SMC'' models cannot
reproduce the data in Figure \ref{COLORCOLOR}. \citet{BOT04} and
\citet{BERNARD08} showed that a similar case holds for the \citet{DESERT90}
models. The main stumbling block is reproducing the observed 60$\mu$m
\citep{DESERT90} or 70$\mu$m \citep{DRAINE07A} emission.

Equation \ref{POLYFIT} is not a unique description. A simple alternative is a
modified blackbody with twice the expected emission at $70\mu$m. In this case:

\begin{equation}
\label{FRAC70}
\frac{I_{\rm 70}}{I_{\rm 160}} = 2.0 \times \left(\frac{70}{160}\right)^{-1.5}~
\frac{B_{\nu} \left(T_{\rm dust}, 70\mu{\rm m}\right)}{B_{\nu} \left(T_{\rm dust},160\mu{\rm m} \right)}~,
\end{equation}

\noindent This is shown by the dash-dotted line in Figure \ref{COLORCOLOR}. It
reproduces the data near N83 with about the same accuracy as Equation
\ref{POLYFIT}. If equilibrium emission sets $I_{\rm 100}/I_{\rm 160}$, then
Equation \ref{FRAC70} implies that other processes (e.g., single-photon
heating of small grains) contribute $\approx 50\%$ of the emission at 70$\mu$m
near N83 (and across the whole SMC). This is in reasonable agreement with the
results for the Solar Neighborhood and several nearby GMCs
\citep{DESERT90,SCHNEE05,SCHNEE08}.

The aim of this paper is not to investigate the details of small grain heating
in the SMC, so we move forward using our empirical fit (Equation
\ref{POLYFIT}). This appears as a solid line in Figure \ref{COLORCOLOR}. It is
a good match to the data near N83, where the RMS scatter in the color of
individual pixels about the fit is $\approx 0.04$. In deriving uncertainties
we use Equation \ref{FRAC70} as an equally valid alternative to Equation
\ref{POLYFIT}.

To convert from $I_{100}/I_{160}$ to $T_{\rm dust}$ we assume that the
SED along each line of sight is described by a modified blackbody with
$\tau_{\lambda} \propto \lambda^{-\beta}$. At long wavelengths
($\lambda \gtrsim 100\mu$m), a blackbody spectrum with a
wavelength--dependent emissivity is indeed a good description of the
integrated SED of the SMC \citep[][]{AGUIRRE03,WILKE04,LEROY07}. We
take $\beta = 1.5$, which is intermediate in the range of plausible
values \citep[e.g.,][]{DRAINE84} and a reasonable description of the
integrated SMC SED from $\lambda \sim 100$--$1000\mu$m. This is not
strongly preferred, and so we allow $\beta$ from $1.0$ to $2.0$ in our
assessment of uncertainties.

\subsection{Uncertainties in  $\tau_{160}$}
\label{RANDOM_UNC}

We assess the uncertainty in $\tau_{160}$ by repeatedly adding realistic noise
to our 70 and 160$\mu$m data and then deriving $\tau_{160}$ under varying
assumptions. For each realization, we offset the observed 70 and 160$\mu$m
maps by a random amount to reflect uncertainty in the background subtraction;
these offsets are drawn from normal distributions with $1\sigma =
0.25$~MJy~ster$^{-1}$ at 70$\mu$m and 1~MJy~ster$^{-1}$ at 160$\mu$m. We add
normally distributed noise to each map. This noise has amplitude equal to the
measured noise (\S \ref{DATA}) and is correlated on scales of $36\arcsec$.

We derive $I_{\rm 100}/I_{\rm 160}$ for each realization using either the
polynomial fit (Equation \ref{POLYFIT}) or scaling the 70$\mu$m intensity
(Equation \ref{FRAC70}), with equal probability of each.  We add normally
distributed noise to $I_{\rm 100}/I_{\rm 160}$ with $1\sigma = 0.04$ (the RMS
residual about Equations \ref{POLYFIT} and \ref{FRAC70}) and then derive
$T_{\rm dust}$ assuming $\beta$ anywhere from 1.0 to 2.0 with equal
probability.

This entire process is repeated 1,000 times. We use the distribution of Monte
Carlo $\tau_{160}$s for each pixel to estimate a realistic uncertainty,
finding individual measurements to be uncertain by $\approx 40\%$
($1\sigma$). We extend the same approach through our derivation of
$N\left(\htwofir \right)$ in \S \ref{N83FIRH2}. In Appendix
\ref{SYSTEMATIC_UNC} we discuss systematic effects that cannot be
straightforwardly incorporated into this approach, two of which (blended dust
populations and hidden cold dust) could impact $\tau_{\rm 160}$.

\subsection{$\tau_{160}$ and Extinction}

It will be useful to make an approximate assessment of the dust column in
terms of $V$-band line-of-sight extinction, $A_V$, and reddening, $E(B-V)$. In
the Solar Neighborhood, $E(B-V) = N\left({\rm H} \right) / 5.8 \times
10^{21}~{\rm cm}^{-2}$ \citep{BOHLIN78} and $\tau_{160} = 2.44 \times
10^{-25}~{\rm cm}^2~N\left(\hi \right)$ \citep[][studying the Galactic cirrus
  where we may safely assume that $N({\rm H}) \approx N(\hi
  )$]{BOULANGER96}. Then

\begin{equation}
\label{EBMVEQ}
\nonumber E(B-V) \left[ {\rm mag}\right] \approx 710~\tau_{\rm 160}~.
\end{equation}

\noindent The reddening law in the SMC yields $R_V \approx 2.7$
\citep{BOUCHET85,GORDON03}, so that

\begin{equation}
\label{AVEQ}
A_{\rm V}\left[ {\rm mag}\right] = 1910~\tau_{\rm 160} \\
\end{equation}

\noindent These equations assume the emissivity, $\tau_{\rm 160}/E(B-V)$, of
Galactic \hi\ but do not depend on the specific dust-to-gas ratio.

Estimates of $A_V$ and $E(B-V)$ based on $\tau_{160}$ and Equations
\ref{EBMVEQ} and \ref{AVEQ} agree well with optical- and UV-based
measurements. \citet{CAPLAN96} compiled $A_V$ for a number of SMC
\hii\ regions, including N83 and N84A/B (both of which lie within the SEST
field). Towards N83 they find $A_V$ in the range $0.42$--$0.79$~mag (mean
$0.63$~mag); towards N84A/B they found $A_V$ from $0.24$--$0.60$~mag (mean
$0.37$~mag). Using their positions and aperture sizes, we derive $A_V = 1.34
\pm 0.36$~mag and $0.93 \pm 0.26$~mag for the same regions. The optical and UV
measurements are based on absorption toward sources inside the SMC. Therefore
they will sample half the total line-of-sight extinction on
average. Accounting for this, our FIR-based extinction estimates are in
excellent agreement with optical values. We find the same good agreement for
Sk~159, a B star near N83 towards which \citet{FITZPATRICK84} and
\citet{TUMLINSON02} measured $E(B-V) \approx 0.05$~mag, while we estimate
$E(B-V) = 0.08 \pm 0.03$~mag (see \S \ref{N83DGR}).

\section{Dust and Gas Near N83}
\label{GASANDDUST}

Following the method described in \S \ref{DUST}, we calculate
$\tau_{160}$ over every line of sight in a $2\arcdeg$ field centered on N83
(Figure \ref{TWODEGREEFIELD}) and in the SEST field. In the process, we derive
a median $T_{\rm dust} = 20.9 \pm 1.5$. This agrees with the $T = 22\pm 2$~K
found by \citet{BOT04} for dust in the SMC Wing. The temperature in the N83
complex is somewhat higher, with median $T_{\rm dust} = 22.9 \pm 1.5$~K and
values up to $\sim 28 \pm 2$~K. The hottest regions are coincident with the
N83, N84A, and N84B \ion{H}{2} regions.

Our goal in this section is to combine $\tau_{160}$ with the measured $N(\hi
)$ to estimate $N(\htwo )$ via

\begin{equation}
\label{NH2EQ}
N (\htwofir ) = \frac{1}{2} \left( \frac{\tau_{160}}{DGR} - N( \hi ) \right)~.
\end{equation}

\noindent Here $DGR$ is the dust-to-gas ratio defined by

\begin{equation}
\label{DGREQ}
\tau_{\rm 160} = DGR~N({\rm H} )~\left[{\rm cm}^{-2}\right]~,
\end{equation}

\noindent $N({\rm H} ) = N(\hi ) + N(\htwo )$, and \htwofir\ refers to the
distribution of \htwo\ derived using this approach. To calculate \htwofir , we
first compare $\tau_{160}$ and $N\left( \hi \right)$ in the area around N83
where the ISM is likely to be mostly \hi\ (\S \ref{HIVSFIR}). This demonstrates
that $\tau_{160}$ effectively traces the ISM and allows us to directly measure
$DGR$ in the diffuse ISM. We show that residuals about this
$\tau_{160}$-$N\left( \hi \right)$ relation come exclusively from regions of
active star formation (\S \ref{RESIDUALS}). We then adopt a reasonable value
for the $DGR$ in N83 itself and estimate $N\left( \htwo \right)$ across the
complex.

\subsection{\hi\ and Dust Near N83}
\label{HIVSFIR}

\begin{figure}
\plotone{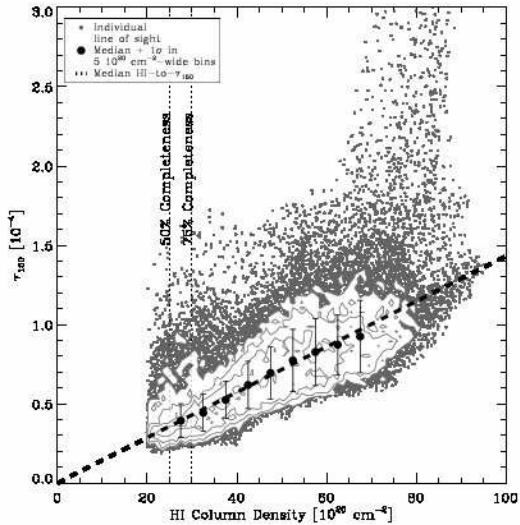}
\caption{\label{FIRVSHI} Dust column, traced by $\tau_{160}$, ($y$-axis) as a
  function of \hi\ column density ($x$-axis) in a $2\arcdeg$ field centered on
  N83. Black circles show average $\tau_{160}$ and $1\sigma$ variation in bins
  $5 \times 10^{20}$~cm$^{-2}$ wide. The dashed line shows the median ratio
  $\tau_{160} / N(\hi ) = 1.4 \times 10^{-26}$~cm$^2$. Dotted lines show the
  {\sc Hi} column density for which $50$ and $75\%$ of the pixels are well
  above the background (see \S 2.1; only such points are plotted). Contours
  show the distribution of data where point density is very high; the contour
  increment is a factor of two in data density. Dust and gas are reasonably
  related by a linear scaling over most of the field. The deviations to high
  $\tau_{160}$ mostly coincide with N83 and other sites of active star
  formation.}
\end{figure}

%, which is known to harbor a substantial reservoir of molecular gas
%  \citep{BOLATTO03}

In Figure \ref{FIRVSHI}, we plot $\tau_{160}$ as a function of $N(\hi )$ over
the $2\arcdeg$ field centered on N83.  Most of the data are well-described by

\begin{equation}
\label{HIDGREQ}
\tau_{\rm 160} = 1.4^{+0.8}_{-0.5} \times 10^{-26}~{\rm cm}^{2}~N(\hi
)~\left[{\rm cm}^{-2}\right]~,
\end{equation}

\noindent which is shown by the dashed line in Figure \ref{FIRVSHI}. We expect
that $N({\rm H}) \approx N(\hi )$ over most of this area. Thus, the clear,
linear correlation in Figure \ref{FIRVSHI} demonstrates that $\tau_{160}$
traces the ISM well here and the slope is an estimate of the $DGR$ in the
diffuse ISM of the SMC Wing.

Equation \ref{HIDGREQ} is consistent within the uncertainties with results of
\citet{BOT04}, who found $\tau_{\rm 160} \sim \left(1.0 \pm 0.5\right) \times
10^{-26}~{\rm cm}^2~N(\hi )~\left[{\rm cm^{-2}}\right]$ for the whole Wing
(after adjusting for slight differences in $T_{\rm dust}$, $\beta$, and
$\lambda$). In the Solar Neighborhood, $\tau_{160} \approx 2.44 \times
10^{-25}~{\rm cm}^2~N(\hi )~\left[{\rm cm^{-2}}\right]$
\citep{BOULANGER96}. Comparing this to Equation \ref{HIDGREQ} implies that the
$DGR$ near N83 is $17^{+10}_{-6}$ times smaller than the Galactic value. This
agrees within the uncertainties with the $DGR$ found for the SMC Wing by
\citet[][]{LEROY07}, which is $\approx 10^{+10}_{-5}$ lower than
Galactic\footnote{\citet{LEROY07} made no correction for \hi\ opacity. Doing
  so would improve the agreement with the present measurement.}.

From Equations \ref{HIDGREQ} and \ref{EBMVEQ}, we estimate $N({\rm H} ) /
E(B-V) \approx 10^{+6}_{-4} \times 10^{22}$~cm$^{-2}$~mag$^{-1}$. This matches
the SMC--average $N({\rm H} ) / E(B-V) \approx 8.7 \times
10^{22}$~cm$^{-2}$~mag$^{-1}$ measured by \citet{FITZPATRICK85} using IUE and
confirmed by \citet{TUMLINSON02} with FUSE.

\subsection{Residuals About the $\tau_{160}$-\hi\ Relation}
\label{RESIDUALS}

\begin{figure*}
\plottwo{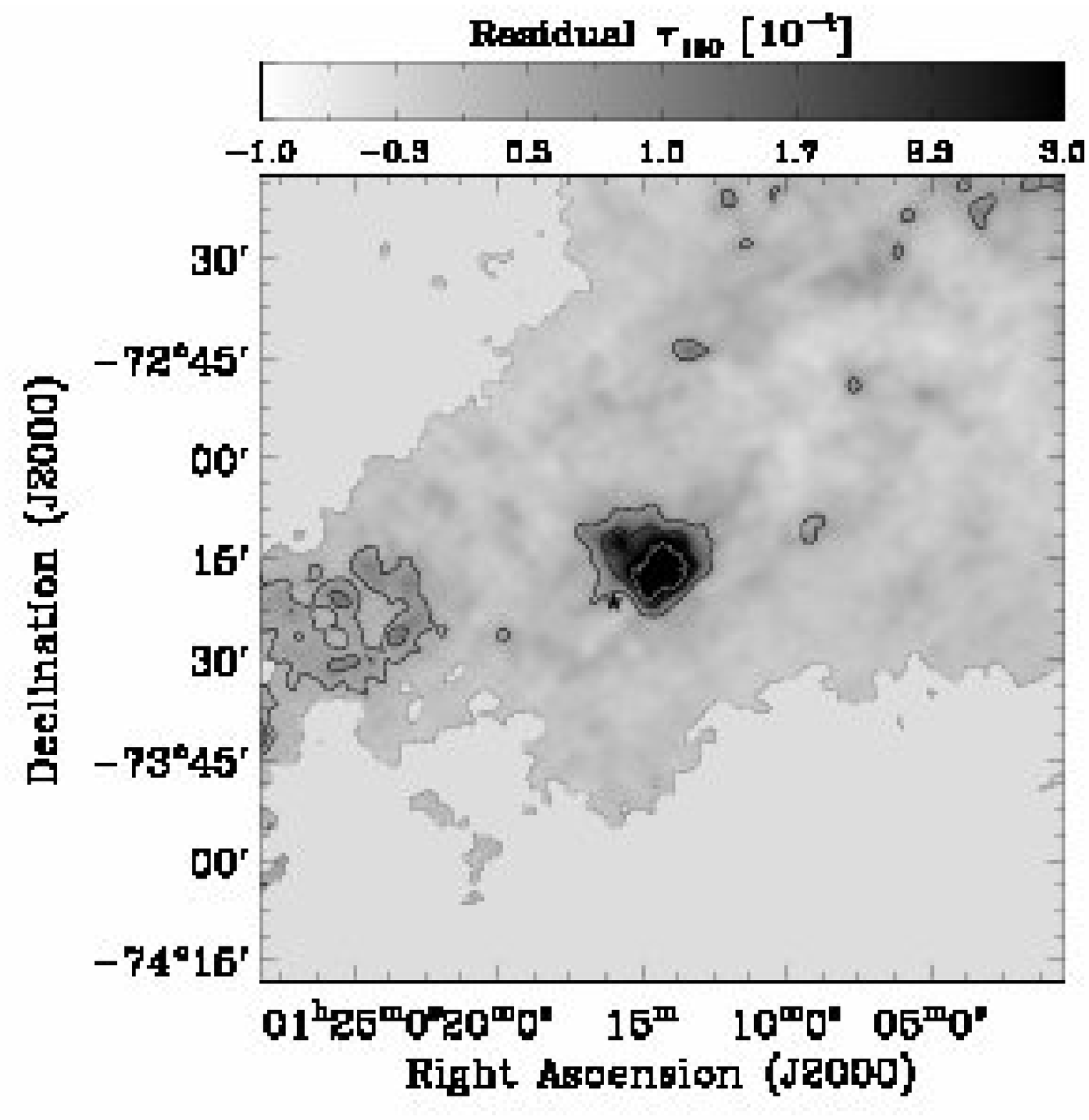}{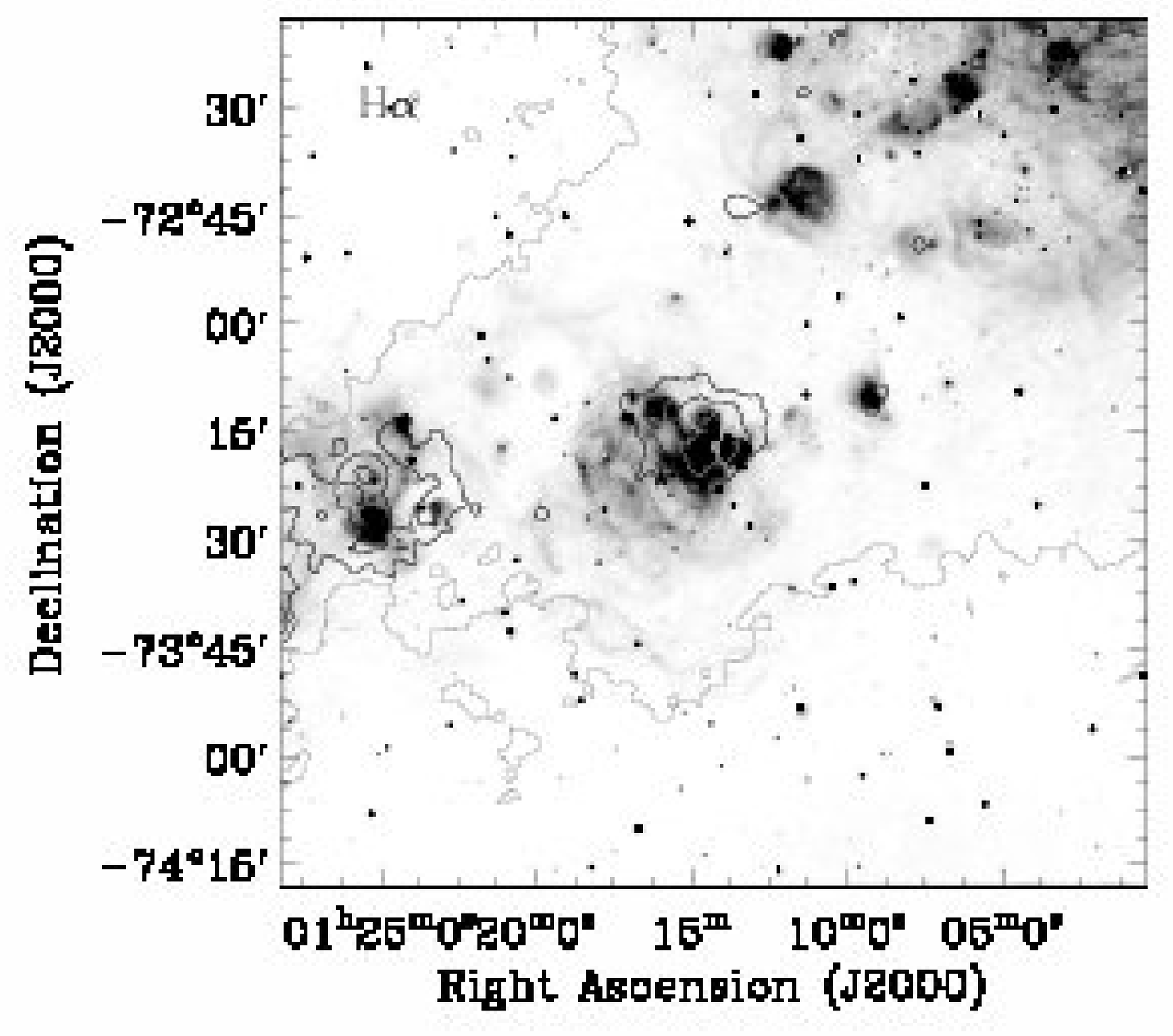}
\caption{\label{TWODEGEXCESS} ({\em left}) Residuals about Equation
  \ref{HIDGREQ}, the average relationship between $\tau_{160}$ and \hi\ in a
  $2\arcdeg$ field centered on N83.  The contours show where our Monte Carlo
  uncertainty estimate yields 85, 98, and 99.9\% confidence that the residual
  is above zero. A thin gray line shows where we clearly distinguish FIR
  emission from the background (see \S \ref{PROCESSING}) and the star
  indicates Sk~159, a B star observed by FUSE and IUE. Most of the region is
  well--described by Equation \ref{HIDGREQ}, but N83 itself shows higher
  $\tau_{160}$ than expected from \hi\ and Equation \ref{HIDGREQ}. ({\em
    right}) The same contours plotted on top of H$\alpha$ emission near N83.
  Regions with high $\tau_{160}$ residuals are associated with sites of recent
  high-mass star formation.}
\end{figure*}

Equation \ref{HIDGREQ} and Figure \ref{FIRVSHI} demonstrate that a single
$DGR$ describes the region near N83 well. The notable exceptions are a small
number of points with high $\tau_{\rm 160}$ relative to their \hi\ column
density. In Figure \ref{TWODEGEXCESS} we show the distribution of residuals
about Equation \ref{HIDGREQ}. Contours indicate where our Monte Carlo
uncertainty estimates yield 85, 98, and 99.9\% confidence that the residuals
are really greater than zero.

The neighboring panel shows the same confidence contours superimposed on an
H$\alpha$ image of the region near N83 (Winkler et al., private
communication). The highest residuals are associated with N83 itself. Other
regions with higher-than-expected $\tau_{\rm 160}$ are also associated with
concentrations of H$\alpha$ emission. H$\alpha$ emission indicates ongoing
massive star formation, which in turn suggests the presence of \htwo . N83
also has significant CO emission, another signpost of H$_2$
\citep{MIZUNO01}. If a large amount of the ISM is \htwo , we expect high
residuals about Equation \ref{HIDGREQ} even for a fixed $DGR$.

\subsection{The Dust-to-Gas Ratio in N83}
\label{N83DGR}

To derive \htwofir\ from Equation \ref{NH2EQ} over the SEST field, we must
know the $DGR$ in N83 itself. We cannot measure this directly because we do
not have an independent measure of the H$_2$ column. We might expect $DGR$ in
N83 to differ somewhat from that in the surrounding diffuse gas of the Wing:
stars are more likely to form in regions with high $DGR$ and the denser
environment may shelter grains from destruction by shocks or lead to grain
growth \citep[e.g.,][]{DWEK98}. In addition to our measurement of the diffuse
ISM, we consider two pieces of evidence when adopting a $DGR$ to use in N83:
observations of a nearby B star and the metallicity of the N84C \hii\ region.

{\em FUSE and IUE Measurements of Sk 159:} From FUSE and IUE absorption
measurements, $E(B-V)$, $N(\htwo )$, and $N(\hi )$ are known towards Sk~159, a
B0.5 star near N83 (marked by a star in Figure \ref{TWODEGEXCESS}). \htwo\ is
detected but the column density is small \citep[$\approx 2 \times
  10^{19}$~cm$^{-2}$,][]{ANDRE04}. The reddening associated with the SMC is
$\approx 0.05$~mag \citep{FITZPATRICK84,TUMLINSON02}, though somewhat
uncertain. The \hi\ column measured from absorption along the same line of
sight is $2\pm1 \times 10^{21}$~cm$^{-2}$ \citep{BOUCHET85}, roughly half of
the column inferred from 21~cm emission along the line of sight (two
kinematically distinct \hi\ components are visible in emission towards Sk~159;
only one of them is seen in absorption, implying that Sk~159 sits between the
two, behind the smaller one). These values imply $N({\rm H} ) / E(B-V) \approx
2$--$6 \times 10^{22}$~cm$^{-2}$~mag$^{-1}$, or $DGR \approx 2$--$7 \times
10^{-26}$~cm$^2$.

{\em Metallicity of N84C:} \citet{RUSSELL90} measured the nebular metallicity
of the N84C \hii\ region, which lies within the SEST field, finding $12+\log
{\rm O/H} = 8.27$, $2$--$3$ times lower than the Solar Neighborhood value and
among the highest for any region the SMC. Translating metallicity into a $DGR$
is not totally straightforward, because the fraction of heavy elements tied up
in dust may vary with environment. For a fixed fraction of heavy elements in
dust, one would expect $DGR \propto Z^{-1}$. Fits to samples of galaxies yield
power law relationships ($DGR \propto Z^\alpha$) with indices in the range
$\alpha = 1$--$2$ \citep[e.g.,][]{LISENFELD98,DRAINE07B}. This would imply
$N({\rm H} ) / E(B-V) \sim 2$--$7 \times 10^{22}$~cm$^{-2}$~mag$^{-1}$ or $DGR
\sim 2$--$7 \times 10^{-26}$~cm$^2$.

{\em \hi\ and $\tau_{160}$:} Equation \ref{HIDGREQ} offers a lower bound on
the $DGR$ --- N83 is extremely unlikely to have a lower $DGR$ than the
surrounding medium ($N({\rm H}) / E (B-V) \approx 10 \times
10^{22}$~cm$^{-2}$~mag$^{-1}$) and from absorption work we know that there is
not a pervasive massive molecular component in the SMC. The magnitude of the
residuals about this equation towards N83 itself also offer a weak upper bound
on the quantity. If we assume $DGR$ much above $3$ times the value in Equation
\ref{HIDGREQ} then some lines of sight inside the SEST field will have
significantly {\em negative} residuals. If the star-forming region itself is
described by a single $DGR$, then it must be roughly bounded by this value,
which translates to $N({\rm H} ) / E(B-V) \sim 3 \times
10^{22}$~cm$^{-2}$~mag$^{-1}$.

{\em Assumed $DGR$ in N83:} The relatively high metallicity and the
measurement towards Sk~159 are balanced against our observations of a very low
$DGR$ in the nearby ISM and the requirement that $\Sigma_{\rm H2}^{\rm FIR}$
not be significantly and systematically negative. The former suggest $N({\rm
  H} ) / E(B-V) \sim 2$--$7 \times 10^{22}$~cm$^{-2}$~mag$^{-1}$, while the
latter yields $N({\rm H} ) / E(B-V) \sim 3$--$10 \times
10^{22}$~cm$^{-2}$~mag$^{-1}$. In the remainder of this paper we adopt assume
that in N83 itself $N({\rm H} ) / E(B-V) \sim 5 \times
10^{22}$~cm$^{-2}$~mag$^{-1}$, which is intermediate in this range. Then

\begin{equation}
\label{N83DGREQ}
\tau_{\rm 160} = 2.8 \times 10^{-26}~{\rm cm}^{2}~N({\rm H} )~\left[{\rm
    cm}^{-2}\right]~.
\end{equation}

\noindent This is twice the value found in the diffuse gas of the SMC Wing
(Equation \ref{HIDGREQ}) and more similar to that found in the actively
star-forming SMC Bar \citep[e.g.,][]{WILKE04,LEROY07}. It is roughly
consistent with observations of Sk~159 and the metallicity of N84C. This $DGR$
also leads to reasonable agreement between dynamical and dust masses in the
star-forming region (\S \ref{DYNAMICS}), which was a factor in settling
on this value. In Appendix \ref{SYSTEMATIC_UNC} we illustrate the effects of
changing this value on our analysis.

\subsection{\htwofir\ in N83}
\label{N83FIRH2}

\begin{figure*}
\plotone{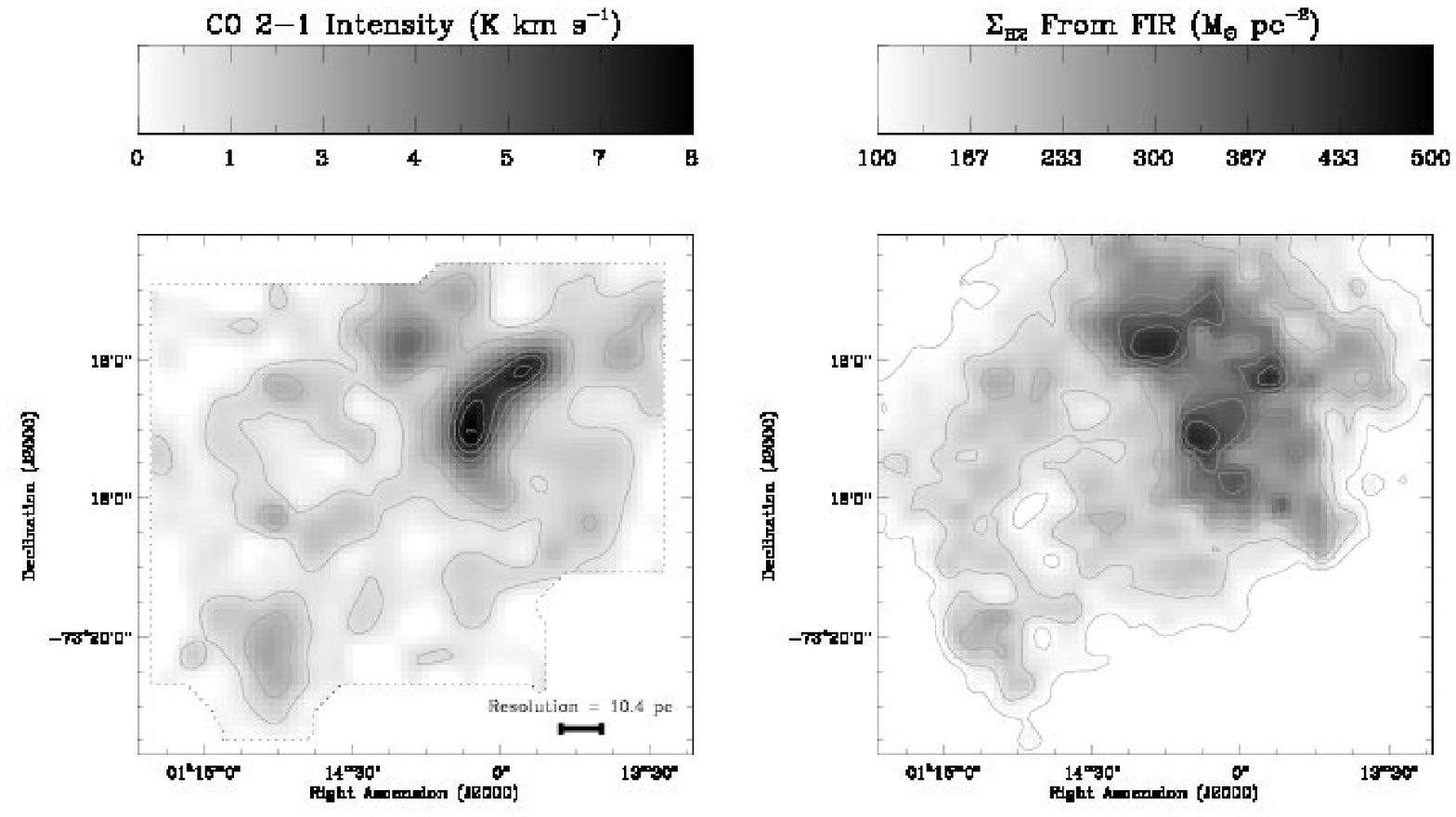}
\plotone{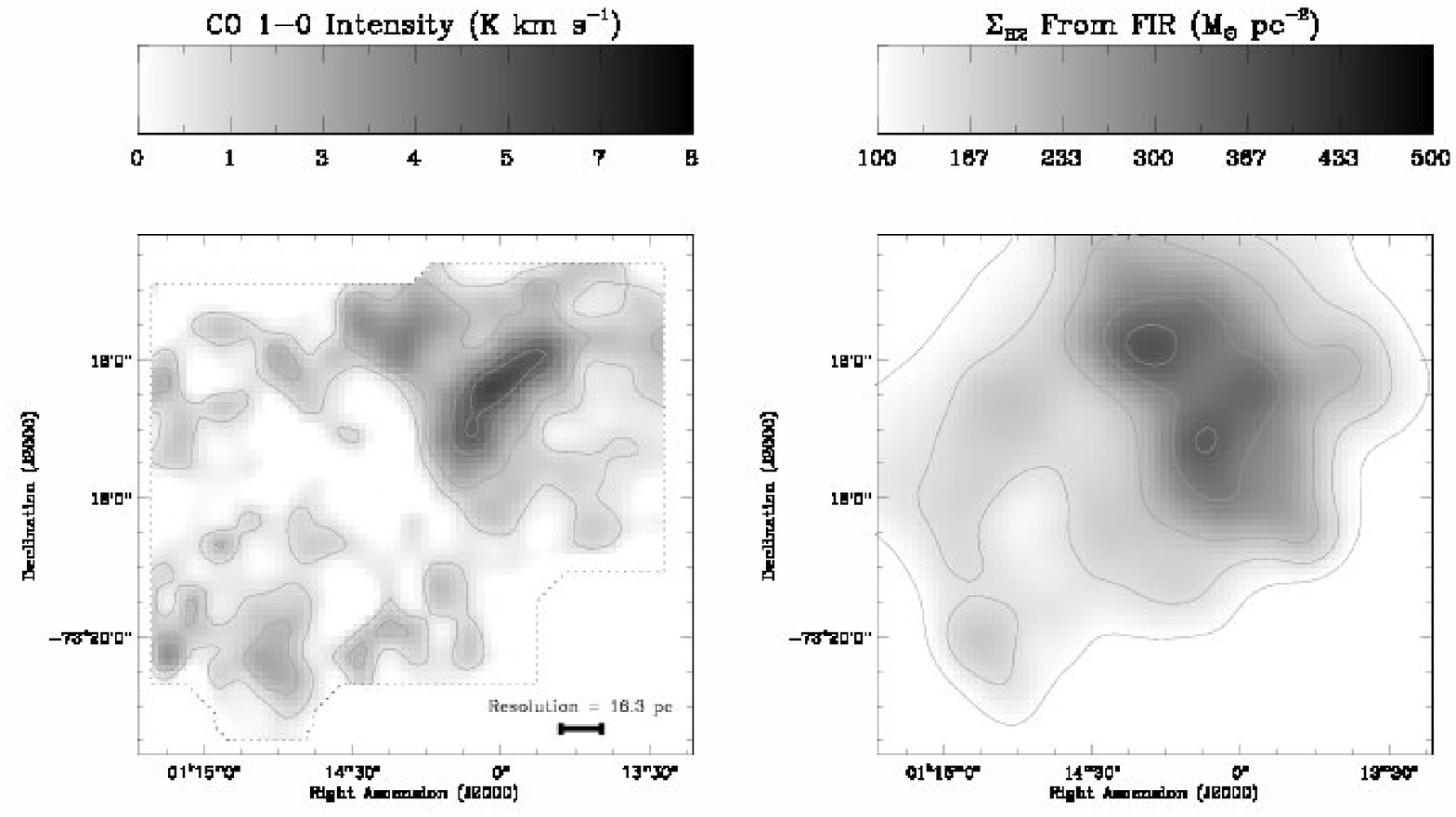}
\caption{\label{HIRESEXCESS}CO emission (left panels) and $\Sigma_{\rm
    H2}^{\rm FIR}$ estimated using Equation \ref{NH2EQ} (right panels). The
  top panels show CO $J=2\rightarrow1$ emission and \htwofir\ at $\sim
  38\arcsec$ resolution (but see \S \ref{RESOLUTION} regarding the resolution
  of the \htwofir\ map). The bottom panels show CO $J=1\rightarrow0$ emission
  and \htwofir\ at $55\arcsec$ resolution. Dotted contours show the boundaries
  of the SEST map.  In the CO maps, contours show $I_{\rm CO}$ from 1 to
  8~K~km s$^{-1}$ spaced by 1~K~km~s$^{-1}$. In the \htwofir\ maps, contours
  indicate $\Sigma_{\rm H2}^{\rm FIR}$ from $100$ to
  $500$~M$_{\odot}$~pc$^{-2}$ spaced by $50$~M$_{\odot}$~pc$^{-2}$.}
\end{figure*}

Combining Equations \ref{NH2EQ} and \ref{N83DGREQ} we estimate $N (\htwofir )$
from $\tau_{160}$ and $N(\hi )$. From $N (\htwofir )$, we calculate the
molecular gas surface density,

\begin{equation}
  \Sigma_{\rm H2}^{\rm FIR}~\left[{\rm M}_\odot~{\rm pc}^{-2}\right] = \frac{N(\htwofir )}{4.6 \times 10^{19}~\left[\mbox{cm}^{-2}\right]}~,
\end{equation}

\noindent which includes a factor of 1.36 to account for helium\footnote{In
  the rest of the paper, $\Sigma_{\rm H2}^{\rm FIR}$ includes this correction
  for helium, while $N(\htwo )$ or $N(\htwofir )$ refer to column density of
  \htwo\ alone} \citep[after][]{WILSON88}. At the same time we estimate the
extinction along each line of sight using Equation \ref{AVEQ}. Carrying out
these calculations, we work with $N (\hi )$ only in average, because the
resolution of the 160 $\mu$m and CO $J=2\rightarrow1$ data are $\sim
38\arcsec$, while that of the \hi\ map is $98\arcsec$ (\S \ref{RESOLUTION}).

The right column in Figure \ref{HIRESEXCESS} shows the resulting maps of
$\Sigma_{\rm H2}^{\rm FIR}$ in N83 at the resolution of the SEST CO
$2\rightarrow1$ (top) and $1\rightarrow0$ (bottom) data. The left column shows
the CO maps. Note that the stretch on the \htwofir\ images runs linearly from
$\Sigma_{\rm H2}^{\rm FIR} = 100$~M$_\odot$~pc$^{-2}$ to
$500$~M$_\odot$~pc$^{-2}$.

Several systematic uncertainties may affect $N\left( \htwofir \right)$ but are
hard to quantify and so not reflected in our Monte Carlo estimate of the
uncertainties. We consider these in Appendix \ref{SYSTEMATIC_UNC}, finding no
strong reason to doubt that Equation \ref{NH2EQ} yields an approximate
estimate of $N(\htwo )$.

\section{\htwofir , CO, Dust, and Dynamics}
\label{STRUCTURE}

\subsection{\htwofir\ and \hi }

Before we consider the relationship between CO, \htwofir , and dust within
N83, we briefly examine the transition from atomic (\hi ) to molecular (\htwo
) gas in the complex. 

\citet{KRUMHOLZ09A} recently considered the transition from \hi\ to \htwo\ in
galaxies. They argue that inside a complex of mixed atomic and molecular gas,
the ratio of \htwo\ to \hi\ along a line of sight ($R_{\rm H2} = \Sigma_{\rm
  H2} / \Sigma_{\rm HI}$) is mainly a function of two factors: total gas
surface density ($\Sigma_{\rm HI} + \Sigma_{\rm H2}$) and metallicity. Their
calculations agree well with a variety of observations, including FIR-based
estimates of $\Sigma_{\rm H2}$ in the SMC at lower resolution.

\begin{figure}
\plotone{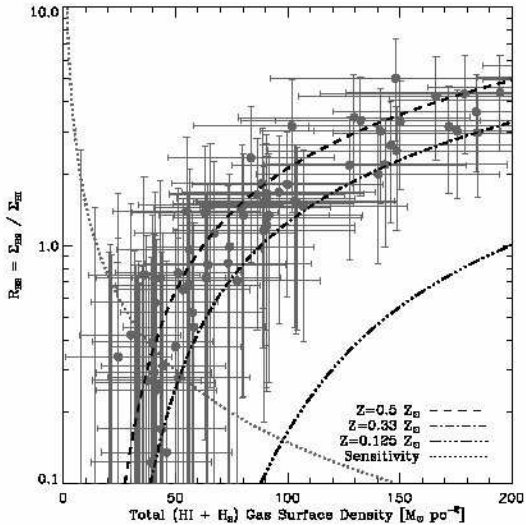}
\caption{\label{RH2VSGAS} Ratio of molecular to atomic gas, $R_{\rm H2} =
  \Sigma_{\rm H2}^{\rm FIR} / \Sigma_{\rm HI}$, as a function of total gas
  column density $\Sigma_{\rm HI} + \Sigma_{\rm H2}$ in the N83 complex (after
  removing \hi\ not associated with the complex). Each point shows an
  independent measurement at $29$~pc resolution. The dotted line shows the
  sensitivity of our \htwofir\ map. We plot the theoretical relationships
  between $R_{\rm H2}$ and $\Sigma_{\rm HI} + \Sigma_{\rm H2}$ calculated by
  \citet{KRUMHOLZ09A} for several metallicities.}
\end{figure}

Comparing \hi\ and \htwo\ in the area around N83, we indeed observe a clear
relationship between $R_{\rm H2}$ and the total gas surface density. We show
this in Figure \ref{RH2VSGAS}, plotting $R_{\rm H2}$ against $\Sigma_{\rm HI}
+ \Sigma_{\rm H2}^{\rm FIR}$ over the whole area where $\Sigma_{\rm H2}^{\rm
  FIR} > 0$. We work at the $98\arcsec$ (29~pc) resolution of the \hi\ map,
with each point in the plot showing an independent measurement. For this
analysis, we are interested in the gas associated with the star-forming
complex itself (not unassociated gas in front of and behind it along the line
of sight). To remove \hi\ unassociated with N83 itself from $\Sigma_{\rm HI}$,
we subtract the median $\Sigma_{\rm HI}$ measured over the area shown in
Figure \ref{TWODEGREEFIELD} ($53$~M$_{\odot}$~pc$^{-2}$) from the measured
$\Sigma_{\rm HI}$ before plotting. This is only an issue for \hi ;
\htwofir\ does not extend beyond the N83 complex.

We overplot the relationship between $R_{\rm H2}$ and $\Sigma_{\rm HI} +
\Sigma_{\rm H2}$ predicted by \citet{KRUMHOLZ09A} for three metallicities:
$Z=0.5$, $0.33$, and $0.125$ times solar. Our data are consistent with the
shape of the \citet{KRUMHOLZ09A} calculation. We find $R_{\rm H2} = 1$ at
$\Sigma_{\rm H2} + \Sigma_{\rm HI} = 68 \pm 12$~M$_{\odot}$~pc$^{-2}$, which
agrees well with their calculations for $Z$ 2-3 times lower than the solar
value. This is roughly the metallicity measured for the N84C \hii\ region
\citep{RUSSELL90}. However, it is significantly higher than the $DGR$ that we
adopt (\S \ref{N83DGR}), which is closer to the lower value. Because
\citet{KRUMHOLZ09A} assume a linear scaling between dust opacity and
metallicity when deriving these curves, this means that there remains some
disagreement between our measurements and their results. Nonetheless, there is
good qualitative agreement in the shape of the curve and the fact that in N83
$R_{\rm H2} = 1$ at a significantly higher value of $\Sigma_{\rm HI} +
\Sigma_{\rm H2}$ than in a solar metallicity cloud.

\subsection{CO and \htwofir }
\label{COANDH2}

\begin{figure*}
\plottwo{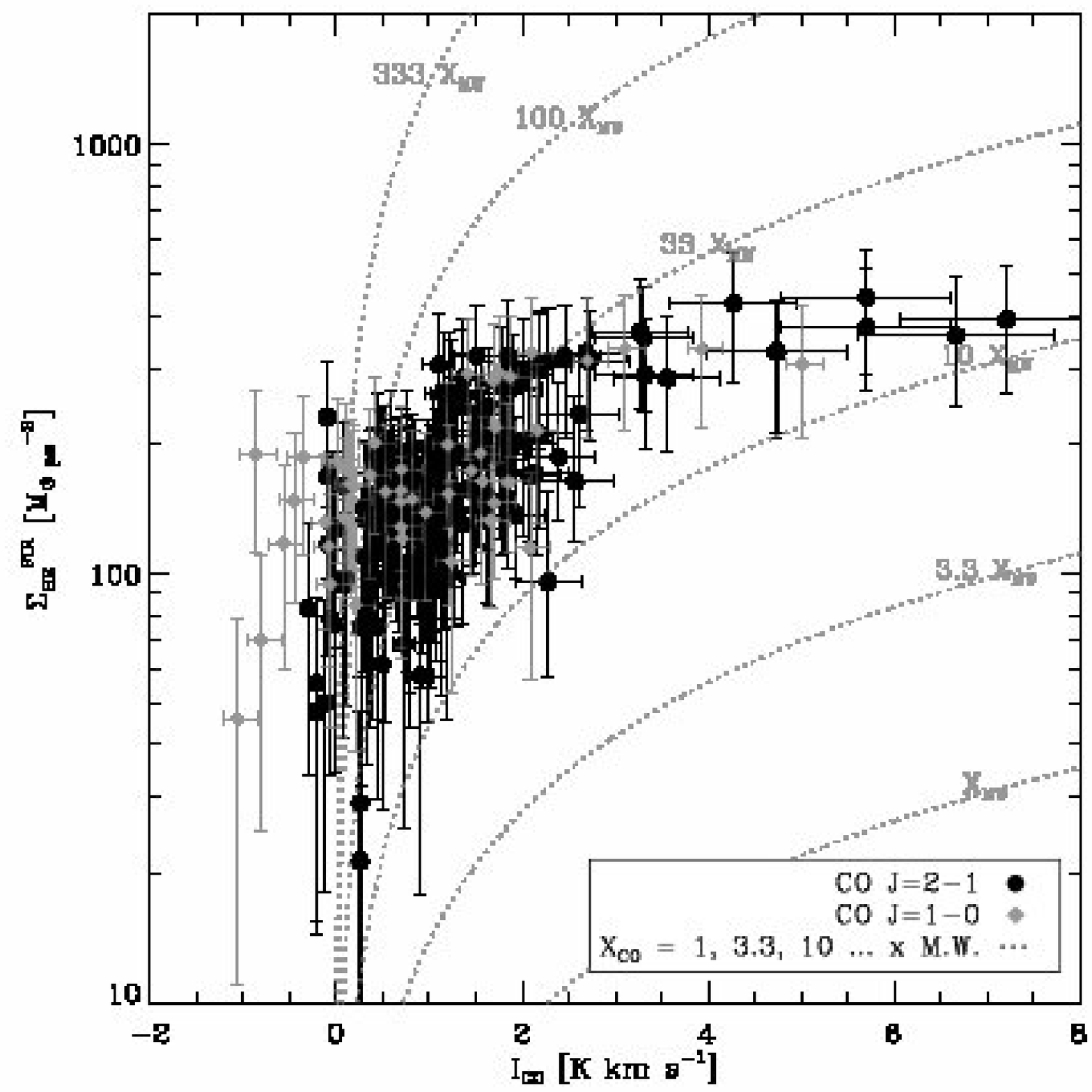}{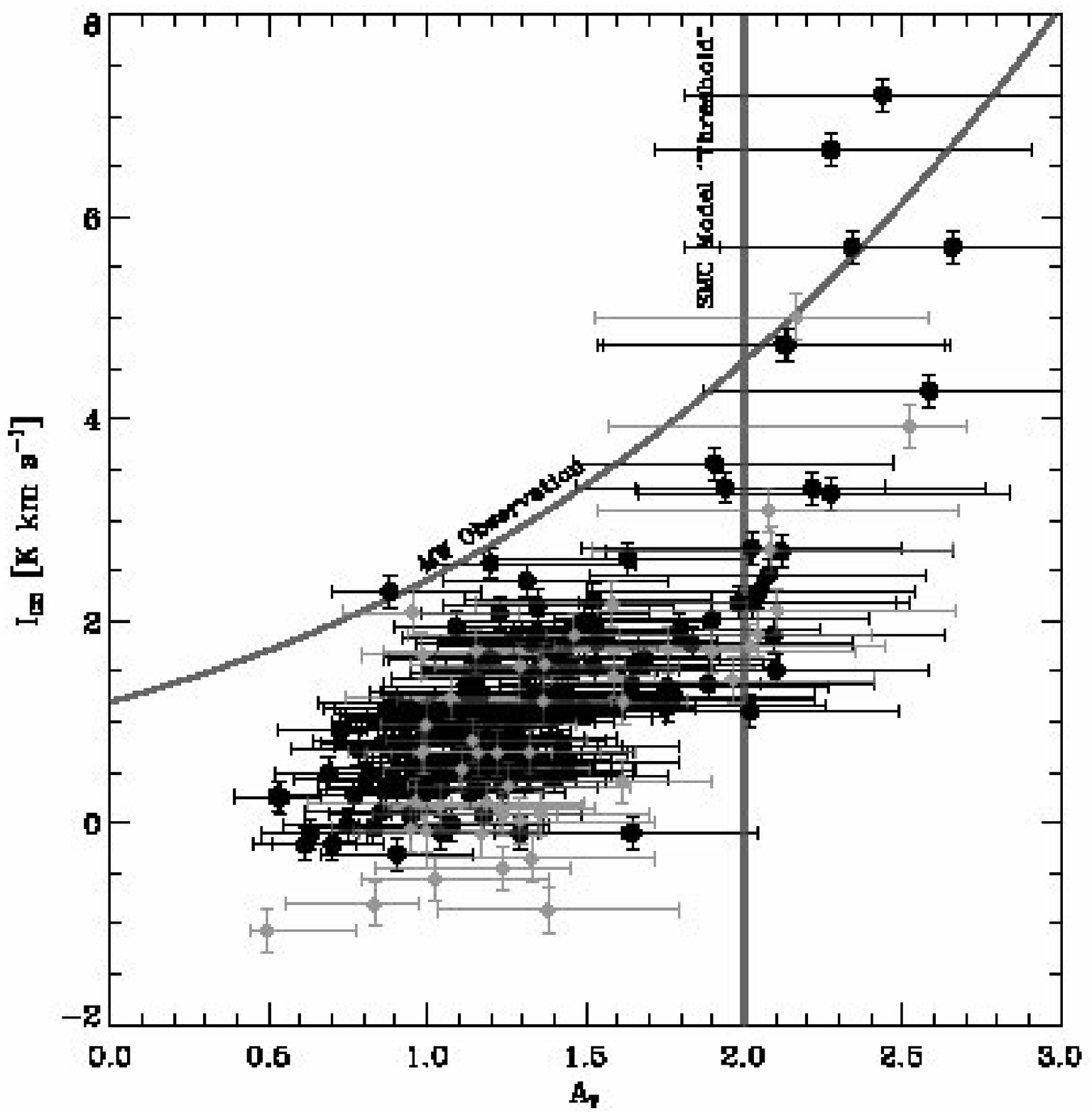}
\caption{\label{COVSH2ANDAV} The structure of H$_2$, CO, and dust in
  N83. ({\em left}) \htwofir\ surface density, $\Sigma_{\rm H2}^{\rm FIR}$, as
  a function CO intensity, $I_{\rm CO}$ ($x$-axis). Dotted gray lines show
  CO-to-H$_2$ conversion factors of 1, 3.33, 10 ... 333 times the Galactic
  value. ({\em right}) $I_{\rm CO}$ ($y$-axis) as a function of line-of-sight
  extinction, $A_V$ ($x$-axis), estimated from $\tau_{160}$. The vertical line
  shows the line-of-sight extinction from which most CO emission emerges in
  models of SMC molecular clouds by \citet{LEQUEUX94}. The gray curve shows
  the relationship between $I_{\rm CO}$ and extinction observed in the Pipe
  Nebula (Milky Way) by \citet{LOMBARDI06}. In both plots, each data point
  represents an independent line of sight.  We show results for the
  $J=2\rightarrow1$ transition ($38\arcsec$ resolution) in black and the
  $J=1\rightarrow0$ transition ($55\arcsec$ resolution) in gray.}
\end{figure*}

Figure \ref{HIRESEXCESS} shows that the distributions of \htwofir\ and CO
share the same peaks and basic morphology. However, the values of $I_{\rm CO}$
in N83 are low compared to a Galactic molecular cloud, which usually show
$I_{\rm CO} \sim 10$~K~km~s$^{_1}$ over a large area, not merely the peaks
\citep[e.g.,][]{WILSON05}. By contrast, the values of $\Sigma_{\rm H2}^{\rm
  FIR}$ (mean $180$~M$_\odot$~pc$^{-2}$) are similar to the surface density of
an average Galactic GMC $\sim 120$--$170$~M$_{\odot}$~pc$^{-2}$
\citep{SOLOMON87,HEYER08}.

This means that CO is faint compared to \htwofir\ in N83. Over the SEST field
\xco\ is

\begin{eqnarray}
  \label{XCOEQ}
  \left< X_{\rm CO}^{2\rightarrow1}\right> &=& 6.7^{+2.8}_{-2.6} \times
  10^{21}~\xcounitsfrac \\ \nonumber \left< X_{\rm CO}^{1\rightarrow0}\right>
  &=& 7.9^{+4.2}_{-2.8} \times 10^{21}~\xcounitsfrac
\end{eqnarray}

\noindent These ratios are $34$ and $40$ times the Galactic conversion factor,
taken to be $\xgal \approx 2 \times 10^{20}$~\xcounits
\citep[e.g.,][]{STRONG96,DAME01}. This value agrees reasonably with previous
FIR-based determinations of \xco\ in the SMC and N83: comparing IRAS and CO at
selected pointings in the SMC, \citet{ISRAEL97} derived $\xco \sim 67~\xgal$.
Applying the same methodology to N83, \citet{BOLATTO03} found $\xco \sim 100
\pm 50~\xgal$. \citet{LEROY07} derived $\xco \sim 50~\xgal$ comparing NANTEN
CO, IRIS 100$\mu$m and {\em Spitzer} 160 $\mu$m towards N83 (removing their
correction for extent).

The left panel in Figure \ref{COVSH2ANDAV} compares \htwofir\ and $I_{\rm CO}$
for individual lines of sight. We plot $\Sigma_{\rm H2}^{\rm FIR}$ as a
function of $I_{\rm CO}$ over the SEST field. We regrid the data so that each
point corresponds to an approximately independent measurement over a $\sim
10$~pc (CO $J=2\rightarrow1$) or $\sim 17$~pc (CO $J=1\rightarrow0$) wide
box. Gray curves show fixed CO-to-H$_2$ conversion factors, starting with
Galactic (lowest) and increasing by factors of 3.33.

As with Figure \ref{HIRESEXCESS}, Figure \ref{COVSH2ANDAV} shows that despite
the very low ratio of CO to \htwofir , the two exhibit an overall
correspondence. High $I_{\rm CO}$ coincides with high $\Sigma_{\rm H2}^{\rm
  FIR}$ and the reverse, so that a rank correlation coefficient of $0.7$
relates the two over the SEST field.

The relationship between $I_{\rm CO}$ and $\Sigma_{\rm H2}^{\rm FIR}$ does not
go through the origin. Instead, $I_{\rm CO} = 0$ corresponds to roughly
$\Sigma_{\rm H2}^{\rm FIR} = 50$--$150$~M$_{\odot}$~pc$^{-2}$. This suggests
the presence of an envelope of \htwofir\ with very little or no associated
CO. Unfortunately, this result is very sensitive to the adopted $DGR$ (\S
\ref{N83DGR} and Appendix \ref{SYSTEMATIC_UNC}). If we take $DGR$ at the upper
end of the plausible range, the data are consistent with no CO-free envelope
although CO emission is still faint relative to $\Sigma_{\rm H2}^{\rm FIR}$ in
the SEST field. If we take $DGR$ at the value derived in the nearby diffuse
ISM, the surface density of the envelope is even higher $\sim
200$-$400$~M$_\odot$~pc$^{-2}$. Although the observation towards Sk~159 does
not actually intersect the envelope in the latter case, it is very nearby and
the low $N(\htwo )$ derived from absorption towards this star offers some
circumstantial evidence against a very massive extended envelope.

The other notable feature of this plot is that at very high $\Sigma_{\rm
  H2}^{\rm FIR}$ CO intensity increases dramatically (the turn to the right at
the top of the plot). We see this in both CO transitions, but the effect is
more pronounced at the higher resolution of the CO $J=2\rightarrow1$ data,
suggesting that the bright CO-emitting structures are still relatively small
compared to the SEST beam. The result is that the line-of-sight integrated
ratio of \htwofir\ to CO is lower for the regions of brightest CO emission,
dropping to $\sim 15$ times the Galactic value. Care must be taken
interpreting these ratios because \htwofir\ and CO emission almost certainly
trace different volumes (\S \ref{DYNAMICS}).

\subsection{CO and Extinction}
\label{COANDAV}

In \S \ref{INTRO}, we highlighted the role of dust in shielding CO from
dissociating radiation. This may provide a simple explanation for the upturn
in CO intensity at high $\Sigma_{\rm H2}^{\rm FIR}$. \citet{LEQUEUX94}
modeled CO emission in SMC molecular clouds.  For their typical cloud ($n_{\rm
  H} \sim 10^4$~cm$^{-3}$, illuminated by a radiation field 10 times the local
interstellar radiation field), they found that most CO emission comes from a
relatively narrow region of the cloud centered on $A_V \sim 1$~mag. Outside
this regime CO intensity is very weak, a scenario that qualitatively matches
what we see in the left panel of Figure \ref{COVSH2ANDAV} \citep[see
  also][]{BELL06}.

In the right panel of Figure \ref{COVSH2ANDAV} we plot CO intensity as a
function of line-of-sight extinction, $A_V$. We estimate $A_V$ from
$\tau_{160}$ using Equation \ref{AVEQ}.  For comparison, we mark $A_V \sim
2$~mag, the line-of-sight extinction that roughly matches the depth from which
\citet{LEQUEUX94} predict most CO emission to emerge (see their Figures 2 and
6). They model a slab illuminated from one side while we estimate the total
extinction along the line of sight through the cloud. Therefore $A_V = 1$~mag
for them corresponds to $A_V \sim 2$~mag for us (though the actual geometry is
likely to be much more complicated). We also plot the relationship between
extinction and CO intensity measured in the Pipe Nebula (a nearby Milky Way
cloud) by \citet[][see their Figure 22]{LOMBARDI06}. We convert $A_K$ into
$A_V$ using their adopted $A_V = A_K/0.112$. They measure a scatter of roughly
$2$~K~km~s$^{-1}$ about this relation.

In agreement with \citet{LEQUEUX94}, we find that lines of sight with bright
CO emission occur almost exclusively above $A_V \sim 2$~mag. Our maps lack the
dynamic range in $A_V$ to test whether $I_{\rm CO}$ is indeed more or less
independent of extinction well above this threshold \citep[as in the Milky
  Way,][]{LOMBARDI06,PINEDA08}. In fact, Figure 3a of \citet{LEQUEUX94} seems
a close match to what we observe: a shallow slope that steepens sharply around
$A_V$ of 2~mag (for us). The radiation field that they assume, $10$ times the
Galactic value is a rough match to what one would infer comparing $T_{\rm
  dust}$ in N83 (median $\sim 23$~K, max $\sim 28$~K) to that of Galactic
cirrus (17.5~K) --- median $\sim5$, maximum $\sim 15$\footnote{For our adopted
  $\beta = 1.5$, the magnitude of the radiation field heating the dust is
  roughly $\propto T^{5.5}$.} --- especially when one recalls that this is
integrated over the whole line of sight rather than tracing the radiation
field incident on the cloud surface.

%, usually by $1$--$2$~K~km~s$^{-1}$

N83 shows somewhat less CO at a given extinction than the Pipe Nebula. This is
also in agreement with the models by \citet{LEQUEUX94}, which predict that CO
from Milky Way clouds emerges from a broader range of $A_V$ and lower values
of $A_V$ than in the SMC. They attribute the difference to lower rates of
photodissociation and it certainly seems likely that the radiation field
incident on the H$_2$ in N83 is much more intense than in the relatively
quiescent Pipe.

Small differences should not overshadow the similarities between the
CO-extinction relation in the Milky Way and that in the SMC. Compared to the
left panel in Figure \ref{COVSH2ANDAV}, the right panel actually shows a
striking similarity between Galactic and SMC clouds. We derive a
CO-\htwo\ conversion that differs with the Milky Way by a factor of $\sim 30$,
while the relationship between extinction and CO is only slightly
offset. Figure \ref{COVSH2ANDAV} supports the hypothesis that shielding,
rather than the distribution of \htwo , determines the location of bright CO
emission. Here ``shielding'' refers to a combination of dust and
self-shielding. Both processes are important to setting the location at which
most C is tied up in CO \citep[e.g.,][]{WOLFIRE93} and the effective shielding
from both sources will be weaker in the SMC than in the Galaxy due to the
decreased metallicity.

Extinction may also be critical to a cloud's ability to form
stars. \citet{MCKEE89} proposed that ionization by an external radiation field
plays an important role in setting cloud structure because it determines the
degree of magnetic support. He predicted that clouds forming low-mass stars in
equilibrium will self-regulate to achieve integrated line-of-sight extinctions
$A_V \approx 4$--$8$~mag. These extinctions are higher than the $A_V \sim
2$--$3$~mag that we find towards the CO peaks N83 or the average extinction
over the region, $A_V \sim 1.5$~mag. We can safely conclude that the N83
region as a whole does not resemble the equilibrium low-mass star forming
cloud described by \citet{MCKEE89}. If these equilibrium structures do exist
in this region, they must be compact relative to our $10$~pc
beam. \citet{BOLATTO08} find that the dynamics of CO emission in the SMC also
appear to disagree with the predictions of \citet{MCKEE89} but present several
important caveats to the comparison. The most important of these here is that
\citet{MCKEE89} explicitly consider clouds forming only low-mass stars, while
N83 is quite obviously actively producing high mass stars.

We emphasize that this comparison between CO and $A_V$ is fairly robust. It
does not depend on our choice of $DGR$, only on the adopted FIR emissivity
($\tau_{\rm FIR} / A_V$) and reddening law. The most likely biases in the
emissivity (e.g., coagulation of small grains) will lower $A_V$, bringing our
results into even closer agreement with those in the Milky Way.

\subsection{\htwofir\ and Dynamical Mass Estimates}
\label{DYNAMICS}

\begin{figure*}
\plottwo{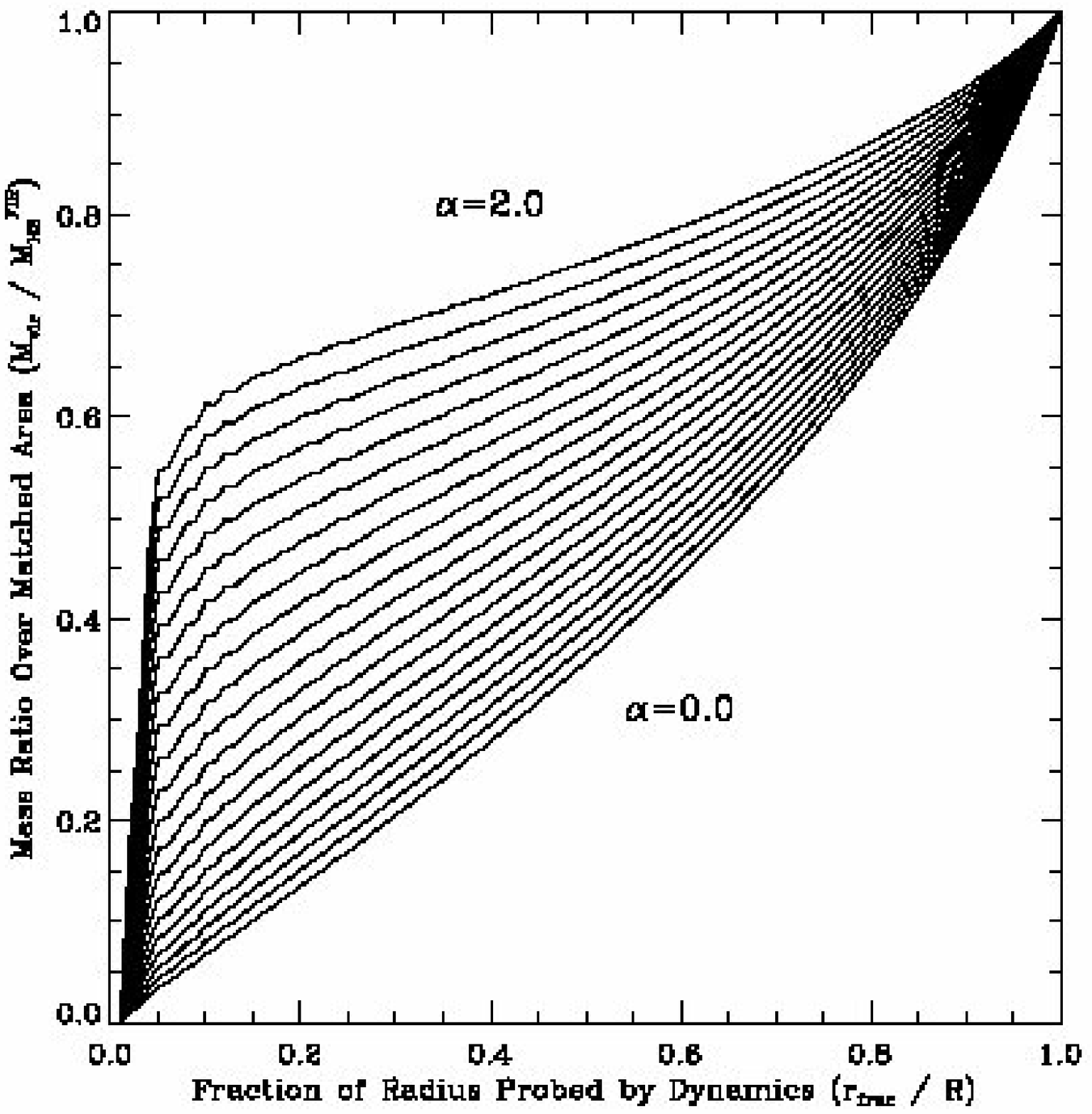}{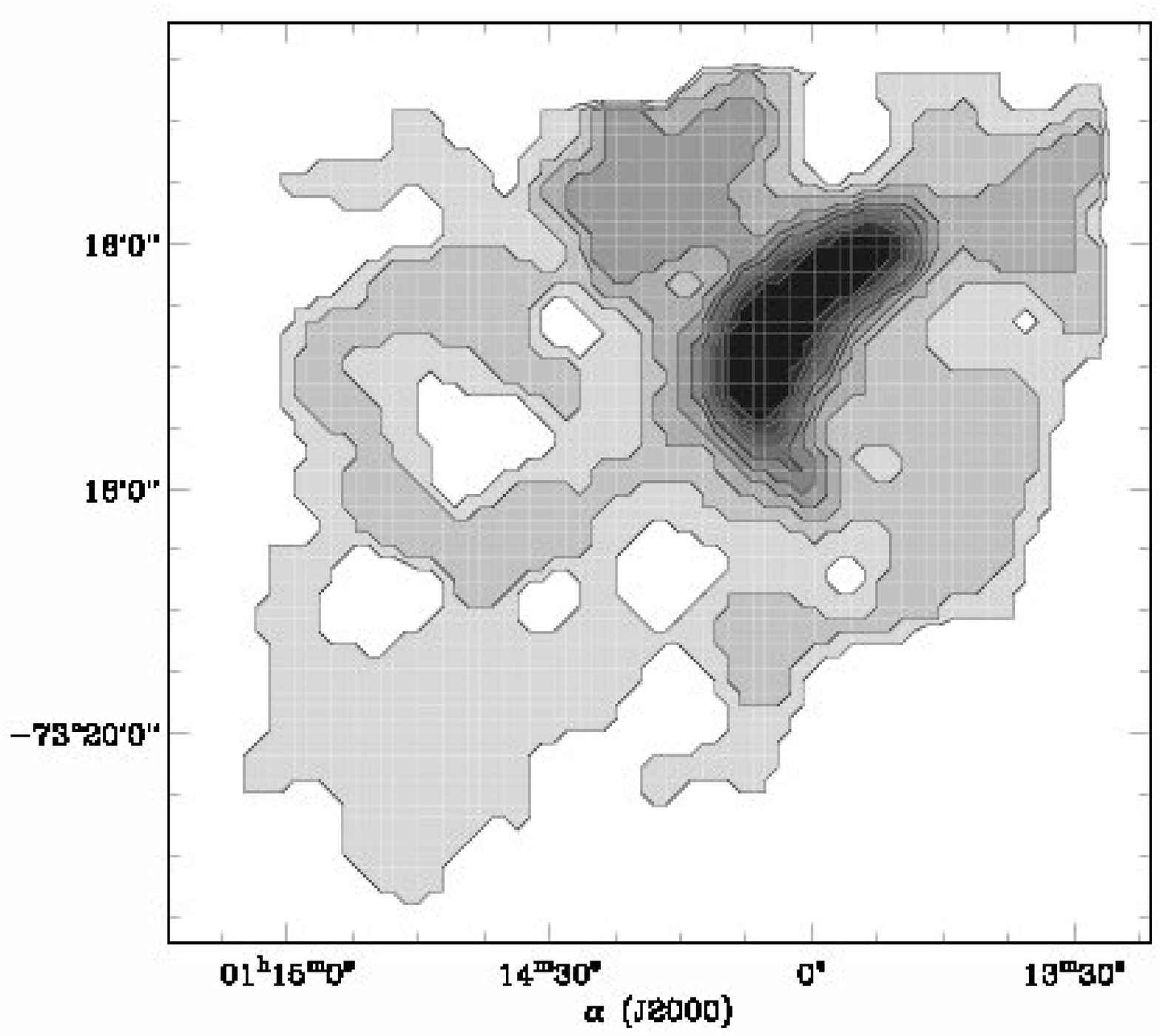}
\plottwo{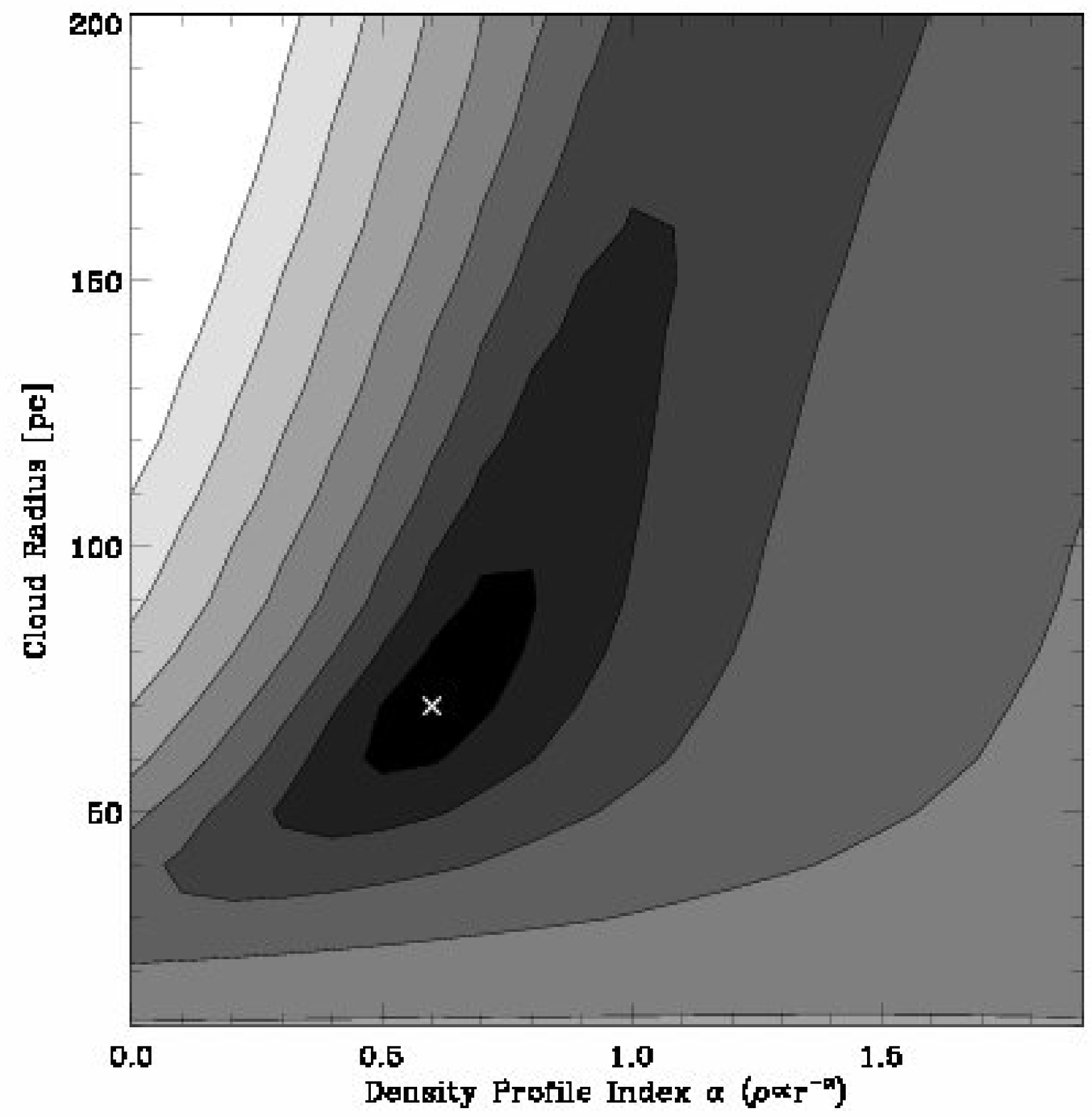}{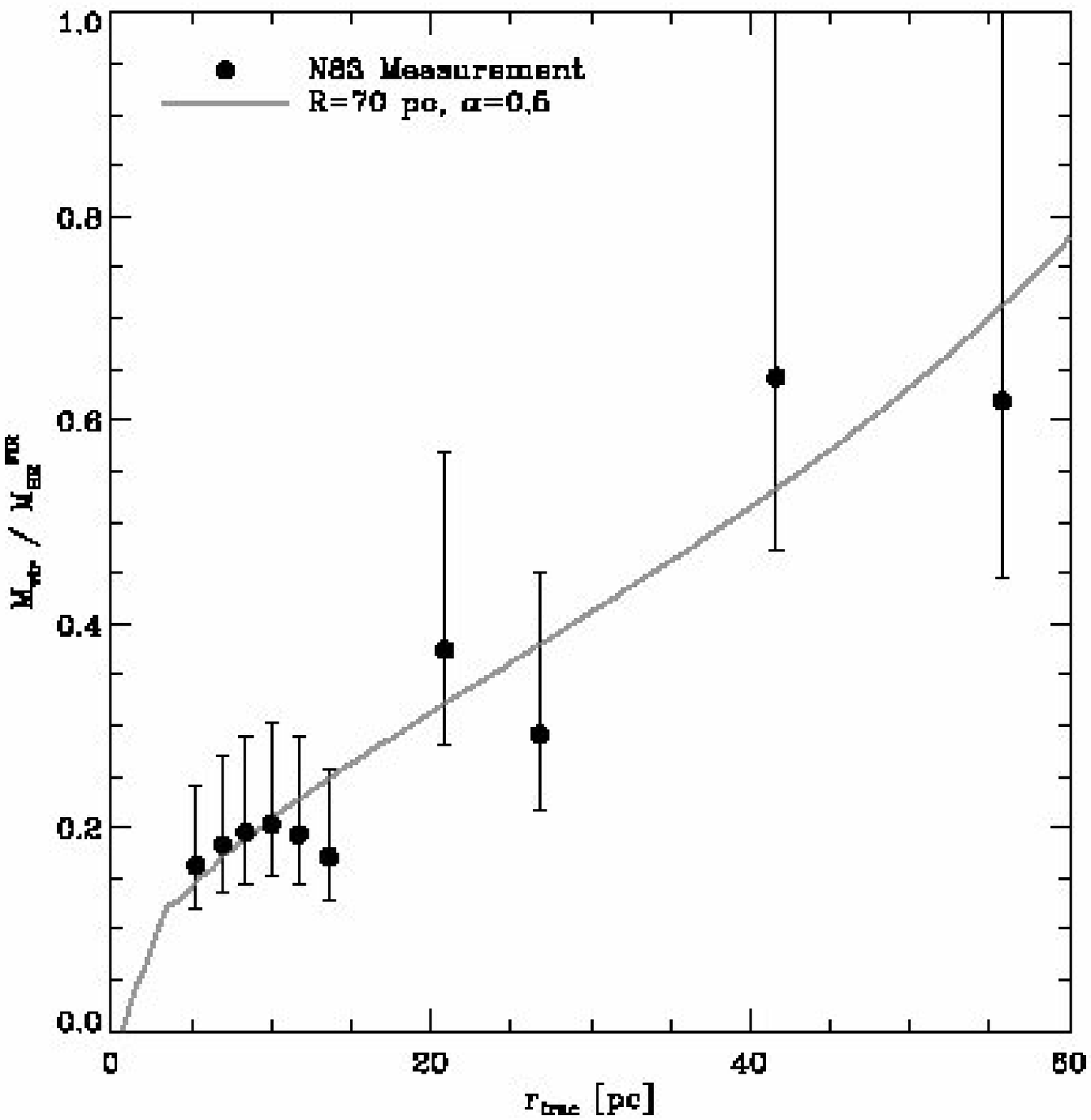}
\caption{\label{DYNAMICSFIG} Reconciling dynamics and \htwofir\ in N83. ({\em
    top left}) The ratio of virial mass, $M_{\rm vir}$, to total H$_2$ mass,
  $M_{\rm H2}^{\rm FIR}$, ($y$-axis) expected for the simple case where
  $M_{\rm vir}$ traces only an inner portion of a cloud (the fraction traced
  is shown on the $x$-axis). Each line shows a cloud with a different density
  profile. ({\em top right}) N83 divided into concentric regions defined by CO
  intensity. We measure $M_{\rm vir}$ and $M_{\rm H2}^{\rm FIR}$ for each
  region. ({\em bottom left}) Results of fitting the models in the top left
  panel to $M_{\rm vir} / M_{\rm H2}^{\rm FIR}$ measured from the regions in
  the top right panel. The $x$-axis show the power law index of the cloud
  density profile; the $y$-axis shows the cloud radius. Contours show reduced
  $\chi^2$, starting at 0.5 and increasing by a factor of $2$ each step. The
  white cross marks the best-fit model ($\rho \propto r^{-0.6}$, $R =
  70$~pc). ({\em bottom right}) $M_{\rm vir} / M_{\rm H2}^{\rm FIR}$ as a
  function of region radius (black points), along with the best fit model
  (gray line).}
\end{figure*}

CO line emission also offers kinematic information. This is the basis of the
virial mass method commonly used to estimate the masses of molecular clouds
and derive CO-to-\htwo\ conversion factors
\citep[e.g.][]{RUBIO93B,WILSON95,ARIMOTO96}, including in N83
\citep{BOLATTO03,ISRAEL03,BOLATTO08}. The potential pitfall of this approach
may be seen from \S \ref{COANDAV}: if CO emission is confined to regions with
extinction above a certain threshold and these regions represent only a
fraction of the whole cloud, then velocity dispersion and size measured from
CO observations will only be lower than their true values. Mass outside the
region of CO emission may exert pressure on the surface of the ``CO cloud,''
but it is not straightforward to estimate the total mass of a cloud from
observing kinematics from only part of it. As a result, in a low-metallicity
cloud like N83, we expect virial masses from CO observations to be smaller
than \htwofir\ (even over matched areas) because the latter also traces the
outer (CO-free) part of the cloud, which exists in front of and behind the
CO-emitting region even over matched lines of sight.

In N83, we have the advantage of an independent measurement of \htwofir\ and
observing the CO emission over a range of scales. Here we test whether these
observations can be reconciled using a simple model in which CO emission comes
from only the inner part of a larger \htwo\ cloud (as appears to be the case
in N83). We consider a spherical cloud with a radially declining density, such
that $\rho \propto r^{-\alpha}$, and a radius $R$ beyond which $\rho = 0$,
i.e., the model usually adopted (with $\alpha = 1$) to calculate cloud virial
masses \citep{SOLOMON87}\footnote{We cap the density at its maximum value over
  the inner $3\%$ of the cloud to avoid divergence.}. We assume that the
dynamical mass estimated from CO line data traces the mass of a fraction of
this cloud, out to radius $r_{\rm frac}$. The ratio of dynamical mass to
$\htwofir$ over a matched area, $M_{\rm vir} / M_{\rm H2}^{\rm FIR}$ is then a
function of $\alpha$ and the ratio of the true radius of the cloud to the
radius of the area being considered, $r_{\rm frac} / R$. The top left panel of
Figure \ref{DYNAMICSFIG} shows this ratio for models with $\alpha$ from 0 to
2.0.

To compare our observations to this model, we measure the line width and
radius of CO emission over a series of scales in N83. We consider intensity
contours in position-position-velocity space, beginning with the bright
northwestern region and including progressively more of the cloud (but always
including that region, see Figure \ref{DYNAMICSFIG}). We estimate the radius
and line width of each region from the area (for the radius) and second moment
(for the line width). To account for the finite resolution of SEST, the radius
of each cloud is adjusted by

\begin{equation}
R = \sqrt{\frac{A_{\rm cloud}}{\pi^{0.5}} - R_{\rm beam}^2}~.
\end{equation}

\noindent Here $A_{\rm cloud}$ is the area of the cloud and $R_{\rm beam} =
0.81 FWHM$ is the ``radius'' of the beam \citep{SOLOMON87}. We combine the RMS
line width, $\sigma_v$, and cloud radius, $R$, to derive the virial mass via

\begin{equation}
\label{VIREQ}
M_{\rm vir} = 1040 R \sigma_v^2 \left[ {\rm M}_{\odot} \right]~,
\end{equation}

\noindent with $\sigma_v$ in km s$^{-1}$ and $R$ in pc. For details of
measuring the properties of extragalactic GMCs from CO emission, we refer the
reader to \citet[][]{ROSOLOWSKY06} and references therein.

For each contour, we measure $M_{\rm vir} / M_{\rm H2}^{\rm FIR}$. We compare
this ratio as a function of $R$ to a range of density profiles and cloud
radii. The resulting distribution of reduced $\chi^2$ is shown in the bottom
left panel of Figure \ref{DYNAMICSFIG}. Our measurements, along with the
best-fit model are shown in the bottom right panel of the same figure.

The best-fit model has $\rho \propto r^{-0.6}$ and $R = 70$~pc, though these
numbers are not strongly constrained. The $\chi^2 = 1$ surface spans $R = 50$
-- $140$~pc and $\alpha = 0.2$ -- $0.8$. Moreover, the assumption of a virial
parameter equal to 1 (i.e., that Equation \ref{VIREQ} holds) is questionable
both because we neglect support by magnetic fields, non-circular geometries,
and surface pressure terms (while considering substructure inside of a larger
cloud). Even more generally, the fundamental assumption that clouds or parts
of clouds are virialized is not certain to hold.

Despite these concerns, Figure \ref{DYNAMICSFIG} does demonstrate that a
simple model --- CO emission nested inside a larger sphere of \htwo --- can
relate dynamics measured from molecular line emission and \htwofir . The best
fit radius, $R = 70$~pc, is quite similar to that needed to achieve the
extinction threshold for CO emission ($A_V \approx 1$) using our adopted $DGR$
and $n \approx 100$~cm$^{-3}$ --- a typical average volume density for
Galactic GMCs and perhaps appropriate for the diffuse gas between dense
molecular clumps in the SMC. These three numbers combine to yield a depth of
$\sim 60$~pc. Meanwhile, the density profile is similar to the $\alpha=1$
commonly used to describe Galactic clouds \citep{SOLOMON87}.

The strong dependence of $M_{\rm vir} / M_{\rm H2}^{\rm FIR}$ on the
size-scale sampled at least partially motivates the discrepancy between
CO-to-\htwo\ conversion factors measured using CO observations and those
derived from dust. At the high resolutions achieved by millimeter-wave
interferometers in Local Group galaxies, CO-emitting clouds are resolved from
their surroundings. By concentrating on these clouds, one samples only dense
regions where CO is well-shielded by dust. This naturally leads to relatively
modest conversion factors. On the other hand, dust measurements and dynamical
measurements made on larger scales sample the whole complex. In the SMC this
appears to includes a large amount of poorly-shielded gas and such methods
therefore return significantly larger conversion factors. One manifestation of
this phenomenon is that dynamical mass determinations from CO measurements
with larger physical beam sizes often return systematically and significantly
higher conversion factors than those obtained from CO measurements in much
smaller beams \citep[][]{RUBIO93A,WILSON95,ISRAEL00,BOLATTO03}. For
interferometer measurements to properly sample the full cloud structure a
multi-scale analysis, such as that presented here or the more rigorous
``dendogram'' approach recently described by \citet{ROSOLOWSKY08}, is
necessary.

Although our dynamical and dust-based results appear consistent with this
simple picture, other recent results suggest a more complex relationship
between the two measurements. Bot et al. (2009, in prep.)  recently measured
the relationship between sub-millimeter dust emission and CO-based dynamical
masses in the southwest part of the SMC Bar. Even after controlling for
contamination by an extended superstructure of CO-free \htwo , they find that
virial masses are systematically lower than dust-based H$_2$ masses on the
scale of individual CO-bright regions. This might arise if clouds are
short-lived (i.e., presently collapsing) or partially supported by magnetic
fields. Alternatively it may reflect altered dust properties in dense cloud
cores. The virial-dust discrepancy measured by Bot et al. and the multiscale
virial-dust measurements presented here can both be readily applied to
simulated clouds and multi-tracer observations of Galactic GMCs. It will be
interesting to see whether these measurements can be replicated purely by
altering the CO-emitting surface inside of a cloud (as it appears from our
simple model) or if they constrain SMC cloud structure to be genuinely
different from that in the Milky Way (as appears to be the case from the Bot
et al. results).

\section{Summary and Discussion}
\label{CONCLUSIONS}

We combine far infrared emission, CO line emission, and a 21-cm \hi\ map to
study the structure of CO, dust, and H$_2$ in the SMC star forming complex
N83.

Two recent surveys of the SMC using {\em Spitzer} \citep[S$^3$MC][and
  SAGE-SMC, Gordon et al., in prep.]{BOLATTO07} allow us to estimate the
distributions of dust and \htwo\ at high spatial resolution. We calibrate a
method to derive the equilibrium dust temperature, $T_{\rm dust}$, and optical
depth at 160$\mu$m, $\tau_{160}$, along the line of sight using only {\em
  Spitzer} data. Applying this method and assuming that the diffuse ISM of the
SMC Wing is mostly \hi , we determine the dust-to-gas ratio ($DGR$) using the
$\tau_{160}$ and \hi\ maps. We find $\tau_{160}$ to be a good tracer of $N(\hi
)$ with $\tau_{160} = 1.4^{+0.8}_{-0.5} \times 10^{-26}$~cm$^2$~$N \left( {\rm
  H} \right)$, implying a $DGR$ $17^{+10}_{-6}$ ($1\sigma$) times lower than
that in the Solar Neighborhood. High residuals about the $\tau_{160}$ --
$N(\hi )$ relation come almost exclusively from regions of active star
formation, with the largest residuals from N83 itself. The most likely origin
for these high residuals is dust associated with \htwo , though several
important systematic uncertainties remain unconstrained (Appendix
\ref{SYSTEMATIC_UNC}). Considering several pieces of evidence (the metallicity
of the N84C \hii\ region, UV spectra of a nearby star, and the $DGR$ in nearby
diffuse ISM) we adopt a $DGR$ of $N({\rm H}) / E(B-V) \approx 5 \times
10^{22}$~cm$^{-2}$~mag$^{-1}$ ($\tau_{160} = 2.8 \times 10^{-26}$~cm$^2$~$N
\left( {\rm H} \right)$) for N83 itself, but note this as a significant
uncertainty with the plausible range spanning $N({\rm H} ) / E(B-V) = 3$--$10
\times 10^{22}$~cm$^{-2}$~mag$^{-1}$. Combining this $DGR$ with $\tau_{160}$
and the measured \hi\ distribution, we derive a map of \htwofir\ in N83.

Comparing CO intensity, kinematics, dust, and \htwo\ we find:

\begin{enumerate}
\item The CO-to-H$_2$ conversion factor averaged over the part of the N83/N84
  region mapped by SEST is very high, $4$--$11 \times 10^{21}$~\xcounits\ or
  $\approx 20$--$55$ times the Galactic value. Despite the large discrepancy
  from the Galactic \xco , there is reasonable agreement between the
  distributions of CO and \htwo\ traced by dust: a rank correlation
  coefficient $\approx 0.7$ relates the two over the SEST field.
\item Bright CO is more confined than \htwo , so that \xco\ varies across the
  region, with the lowest (most nearly Galactic) values near the CO peaks. The
  magnitude (or existence) of an extended, truly CO-free envelope is a
  sensitive function of the adopted $DGR$. Our best estimate is that such an
  envelope does exist, with $\Sigma_{\rm H2} \approx 100$~M$_\odot$~pc$^{-2}$
  where $I_{\rm CO} \sim 0$.
\item CO emission is a function of line-of-sight extinction, which we estimate
  from $\tau_{160}$. Bright CO emission is largely confined to regions with
  $A_V \gtrsim 2$~mag. This agrees well with modeling of SMC clouds by
  \citet{LEQUEUX94} and roughly matches what is seen in the Milky Way. This
  result is robust to most of the systematic uncertainties that affect our
  determination of \htwo .
\item A simple model can reconcile dynamical masses (measured from CO) with
  \htwo\ (measured from dust). In this model, CO emission comes a surface
  within the cloud while dust emission traces all H$_2$ along the line of
  sight. The best-fit density profile and radius are $\rho \propto r^{-0.6}$
  and $R = 70$~pc. These are not strongly constrained, but the density profile
  is similar to that inferred for Galactic clouds and the radius is consistent
  with that required to achieve $A_V \approx 1$ mag for our adopted $DGR$ and
  a typical molecular cloud density.
\end{enumerate}

These results --- particularly the confinement of intense CO to regions of
relatively high line-of-sight extinction --- are all consistent with the
selective photodissociation of CO relative to \htwo\ at low metallicities
\citep[e.g.,][]{MALONEY88,RUBIO93A,RUBIO93B,ISRAEL97,BOLATTO99}. In this
scenario, the distribution of CO emission is largely driven by need for dust
to shield CO from dissociating radiation. The underlying distribution of \htwo
, while subject to significant systematic uncertainties, appears similar to
that in a Galactic GMC complex.

If the distribution of CO emission is indeed largely determined by dust
shielding, then we expect that the ratio of CO emission to H$_2$ mass will
depend sensitively on both the local $DGR$ and the radiation field incident on
the cloud. These effects may largely cancel in more massive spiral galaxies,
yielding a CO-to-H$_2$ conversion factor that is fairly robust
\citep[e.g.,][]{WOLFIRE93}. In low-mass galaxies, which have high radiation
fields and low $DGR$, they will tend to compound, producing extended envelopes
of \htwo\ with little or no associated CO.

From recent large surveys of the Magellanic Clouds at infrared and millimeter
wavelengths \citep[e.g.,][Gordon et al., in
  prep.]{FUKUI99,MIZUNO01,MEIXNER06,BOLATTO07,OTT08}, it will be possible in
the next few years to fill the right panel in Figure \ref{COVSH2ANDAV} with
points from across the Clouds. This will allow the quantification of the
radiation field (and perhaps density) as a ``second parameter'' in the $I_{\rm
  CO}$-$A_V$ relation. It may also allow an improved calibration of \xco\ as a
function of both $DGR$ and local radiation field, extending the pioneering
work by \citet{ISRAEL97} to the scale of individual clouds.

Even with such data, it is unclear if CO emission can remain an effective
tracer of \htwo\ on the scale of individual clouds. Tracing local variations
in $DGR$ and radiation field to apply a spatially variable \xco\ may not be
possible or practical. Of course, CO is already well-known to be a flawed
tracer of \htwo\ {\em within} Galactic clouds \citep[e.g.,][]{PINEDA08} but
retains significant utility for tracing H$_2$ on large scales. Over a sizable
portion of a galaxy, variations in the radiation field and $DGR$ may average
out and allow a calibration to work at a basic level. Given that the options
to trace \htwo\ in low-metallicity galaxies remain limited, a combination of
dust and molecular line emission is likely to be the only widely available
option in the near future. {\em Herschel} spectroscopy of the [CII] line and
{\em Fermi} observations of $\gamma$ ray emission from the Magellanic Clouds,
while both likely to illuminate the issue significantly, will only target a
small sample of galaxies.

\acknowledgements We thank the anonymous referee for a detailed and helpful
critique. We thank Henrik Beuther for helpful comments on a draft of the paper
and Mark Krumholz for a helpful discussion. We acknowledge the use of NASA's
Astrophysics Data System Bibliographic Services.

\bibliography{/a32d6/leroy/bib/akl}

\appendix
\section{Systematic Uncertainties in $N\left(\htwofir \right)$}
\label{SYSTEMATIC_UNC}

\begin{figure*}
  \plottwo{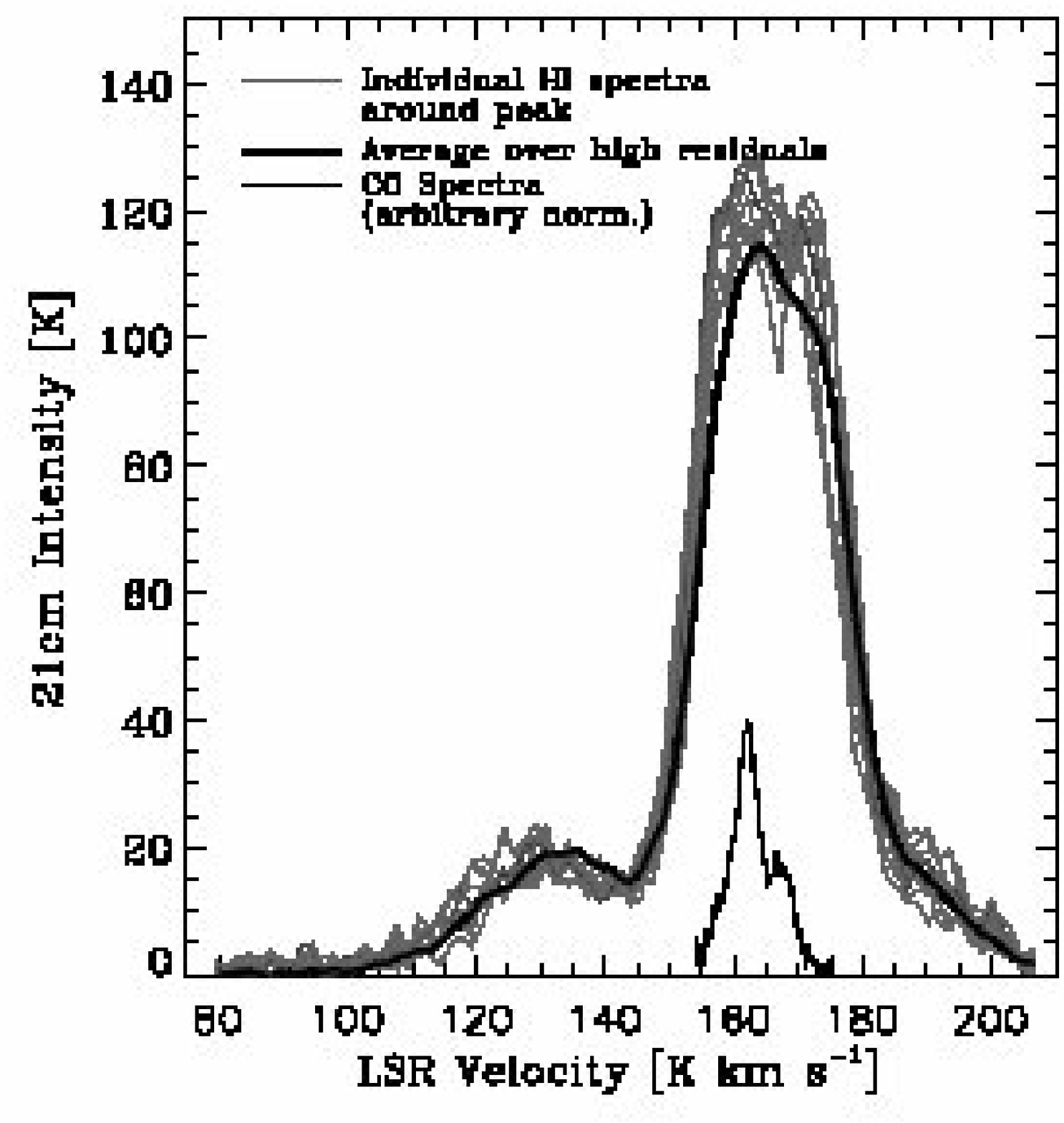}{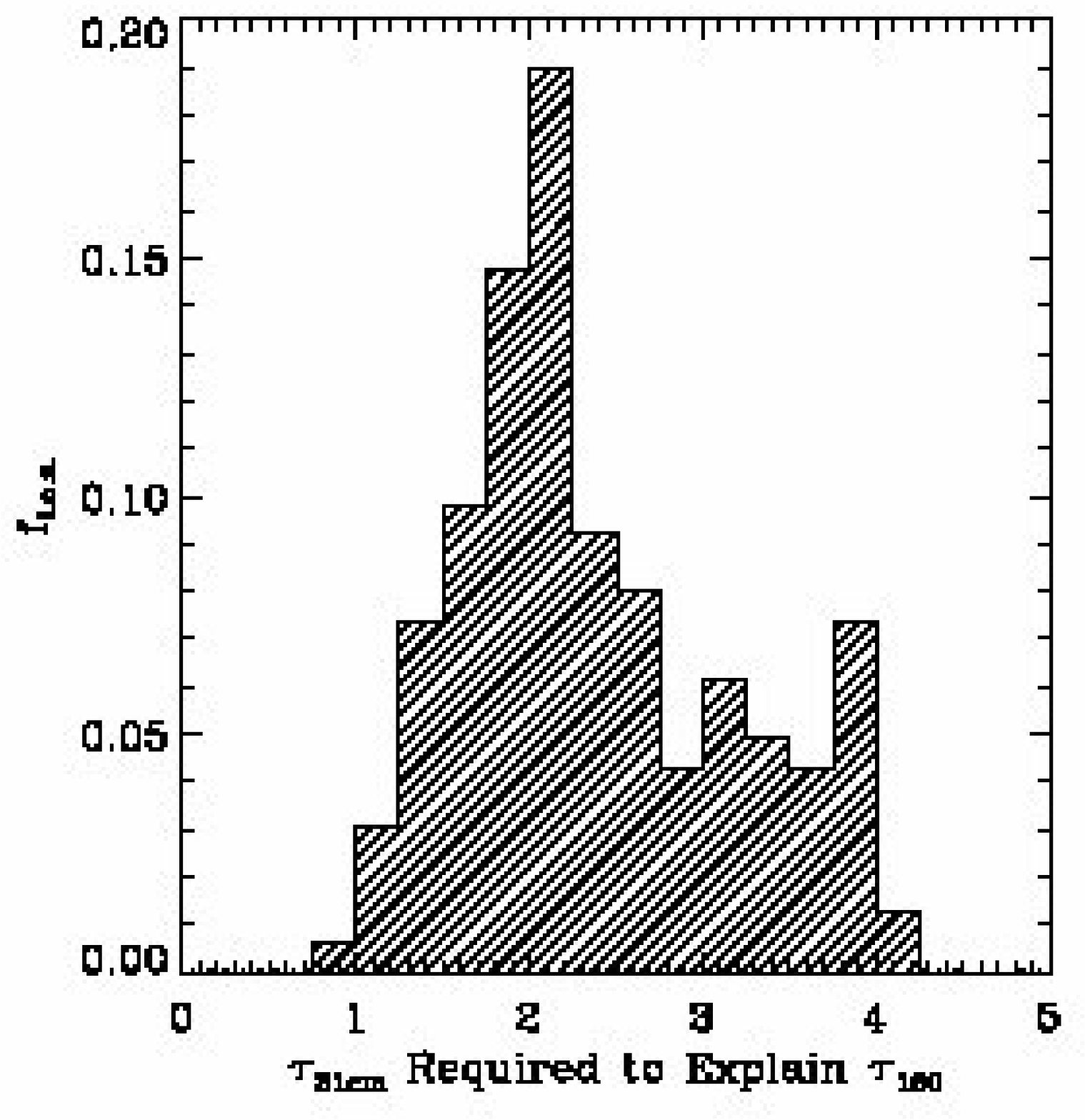}
  \caption{\label{OPAQUEHI} ({\em left}) The average \hi\ spectrum over the
    region of high residuals (black) and spectra from individual lines of
    sight in this area (gray). The spectrum of CO emission (with an arbitrary
    normalization) is shown below the \hi . ({\em right}) The distribution of
    opacities in the (integrated) 21cm line required to explain the residuals
    in highest contour in Figure \ref{TWODEGEXCESS}. Although individual
    spectra show some evidence of optical thickness, we see no clear signature
    of self absorption. The line-integrated values of $\tau_{\rm 21cm}$
    required to explain $\tau_{160}$ in N83 are mostly higher than the peak
    values of $\tau_{\rm 21cm}$ measured anywhere in the SMC by
    \citet{DICKEY00}.}
\end{figure*}

\begin{figure}
  \plottwo{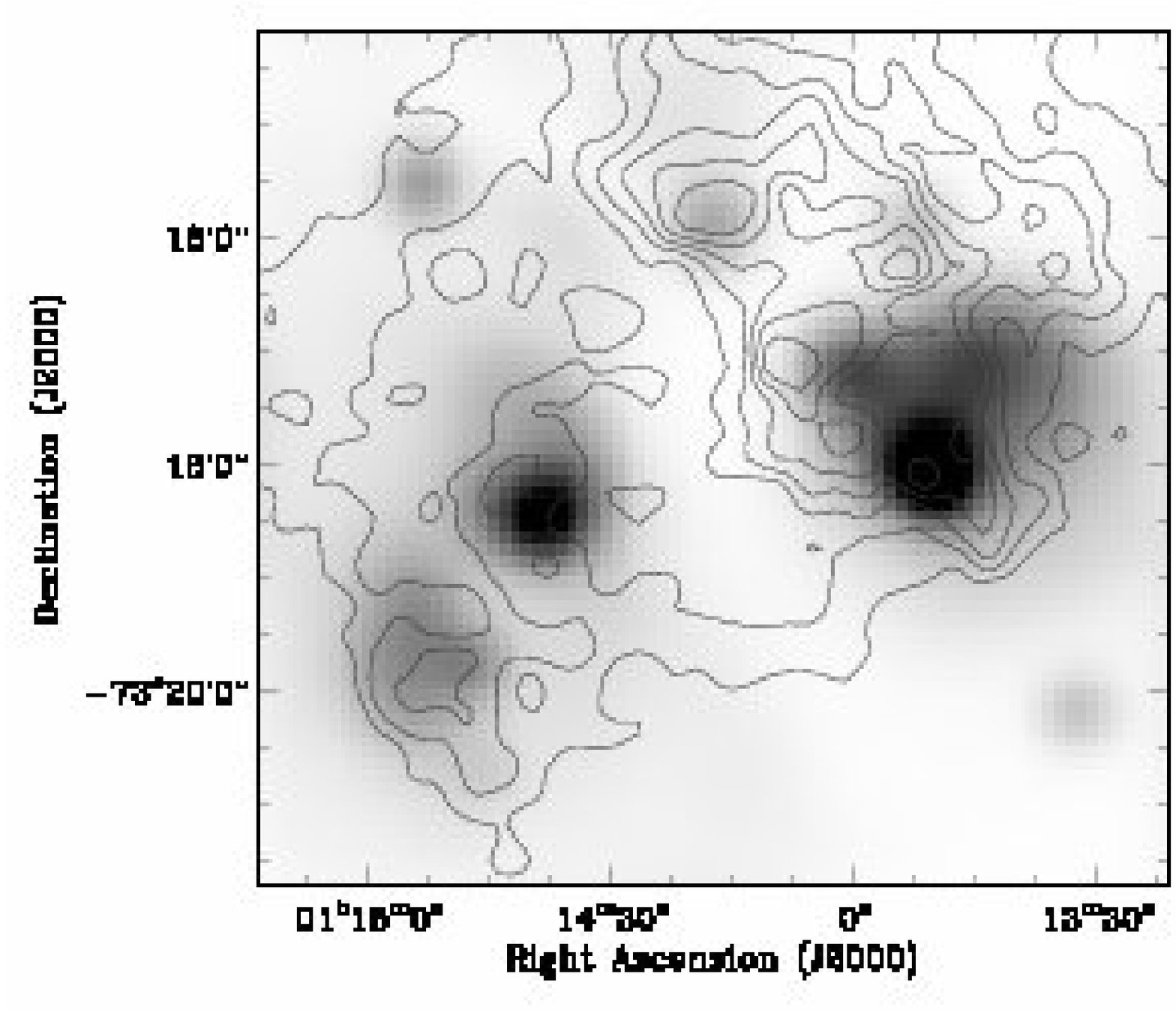}{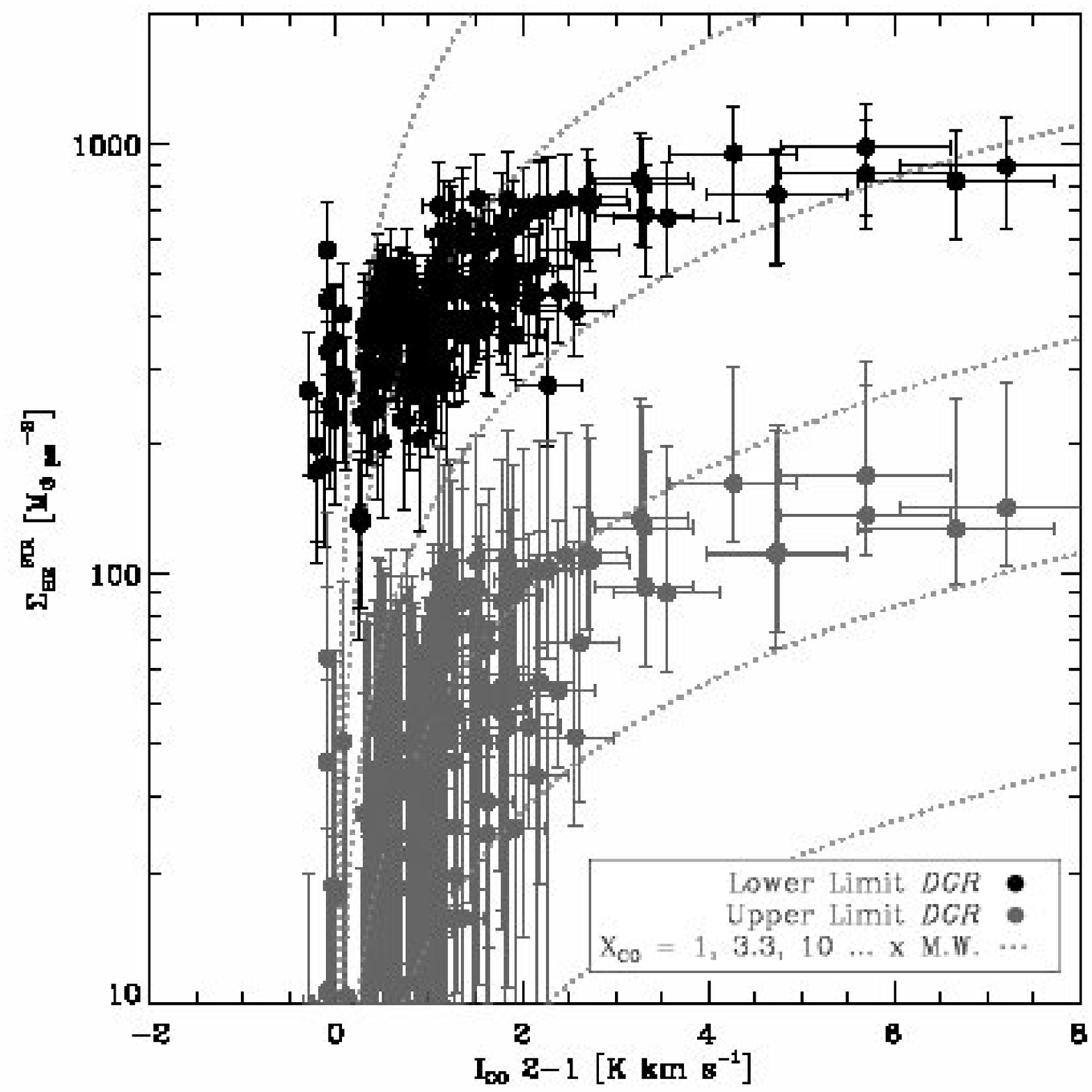}
  \caption{\label{WARMGASANDDGR} ({\em left}) H$\alpha$ emission (gray scale)
    with $\Sigma_{\rm H2}^{\rm FIR}$ shown in contour (both at $36\arcsec$
    resolution). Although H$\alpha$ and \htwofir\ roughly coincide on large
    scales (Figure \ref{TWODEGEXCESS}), the detailed distributions are not a
    good match. ({\em right}) The effect of changing $DGR$ to the most extreme
    plausible values on the relationship between $\Sigma_{\rm H2}^{\rm FIR}$
    ($y$-axis) and $I_{\rm CO}$ ($x$-axis, only $J=2\rightarrow1$ shown). The
    black points show the lowest plausible $DGR$ in N83, that found in the
    nearby diffuse gas. The gray points show the highest plausible $DGR$,
    $\sim 3$ times this value.}
\end{figure}

Several systematic uncertainties may affect $N\left( \htwofir \right)$ but are
hard to quantify and so not reflected in our Monte Carlo estimate of the
uncertainties. Here we discuss these for the specific case of N83 \citep[for a
  more general discussion see][]{ISRAEL97B}. We find no strong reason to doubt
that Equation \ref{NH2EQ} yields an approximate estimate of $N(\htwo )$. N83
appears unlikely to harbor a significant population of cold dust and we do not
observe compelling evidence that dust traces mostly warm ionized gas or high
optical depth \hi . There is likely some blending of populations along the
line of sight, but the magnitude of the effect is unclear. Grain processing is
largely unconstrained, but we note the dissimilarity between N83 and the
dense, cold cores where these effects are usually discussed.

{\em Blending of Populations Along the Line of Sight:} N83 is a dense, active
region and the line-of-sight distance through the SMC may be very long. As a
result, the observed dust emission may represent a blend of several dust
populations with different temperatures. The likely effect is that we
overestimate the average $T_{\rm dust}$ along the line of sight and therefore
underestimate $\tau_{\rm 160}$ and \htwofir\ \citep[e.g., see tests on
  simulated clouds by][]{SCHNEE06}.

{\em Cold Dust:} A related concern is that our longest wavelength data are at
$160\mu$m. As a result, we would miss any population of cold dust. In the
Milky Way, when cold, molecular filaments can be isolated from embedded star
formation, they are often observed to have low dust temperatures ($T \lesssim
15$~K) and little out-of-equilibrium emission
\citep[e.g.,][]{LAUREIJS91,BERNARD99,STEPNIK03}. As with blending of several
populations, cold dust is most likely to be associated with the dense,
molecular environment of N83. Missing cold dust would lead us to underestimate
$\tau_{160}$ and $N(\htwofir )$.

Given the high $T_{\rm dust}$ in N83 and the presence of ongoing, vigorous
star formation we consider it unlikely that there is a significant amount of
cold dust present. We attempt a simple test that reveals the presence of cold
filaments in Galactic GMCs \citep{ABERGEL94,BOULANGER98}: we take the median
$I_{160}/I_{70}$ over the region, scale the $70\mu$m map by this value, and
subtract it from the 160$\mu$m map. This should reveal the location of any
local 160$\mu$m excess, a likely signature of cold dust. We find no such
excess associated with N83 as a whole or the CO peaks in particular.

{\em Other Gas Phases:} We refer to the results of Equation \ref{NH2EQ} as
"\htwofir '' but this is actually an estimate of all gas not traced by the
21-cm transition. Some of this might be high optical depth \hi\ or warm
ionized gas. Neither appears to be a plausible explanation for the majority of
such gas in N83. This agrees with results from the Milky Way, where excess
dust emission identified in a similar way also appears to correspond mostly to
\htwo \citep{DAME01}.

The right panel in Figure \ref{OPAQUEHI} shows the \hi\ opacities required to
account for $\tau_{160}$ in N83 given our adopted $DGR$. These values,
$\tau_{\rm 21cm} = 2$--$4$, are higher than those implied by the fit of
\citet{DICKEY00}, which yields a maximum line-integrated $\tau_{\rm 21cm} \sim
0.55$ (correction factor $\sim 1.3$) near N83. Indeed, most of the
line-integrated values of $\tau_{\rm 21cm}$ in Figure \ref{OPAQUEHI} are
higher than any of the {\em peak} $\tau_{\rm 21cm}$ values (i.e., $\tau_{\rm
  21cm}$ in the most opaque velocity channel) measured by \citet{DICKEY00} in
the SMC (maximum $\sim 1.7$). though that study did not probe any star-forming
peaks; toward the starburst region 30 Doradus in the LMC \citet{DICKEY94}
found peak $\tau_{21cm}$ values of $\sim 2$, which is still too small to
achieve the line-integrated value of $\tau_{21cm}$ required account for
$\tau_{160}$ in N83. The 21cm spectra do show some evidence of optical
thickness at a brightness temperature of $\sim 120$~K, but no clear signs of
self-absorption at the velocity of the CO peak (left panel in Figure
\ref{OPAQUEHI}). We cannot rule out unlucky geometry, but achieving
line-integrated optical depths of $2$--$4$ without invoking a contrived
scenario appears difficult.

Warm ionized gas also seems unlikely to account for most of \htwofir . The
left panel in Figure \ref{WARMGASANDDGR} shows contours of $\Sigma_{\rm
  H2}^{\rm FIR}$ over an H$\alpha$ image (at matched resolution) in the SEST
field. Although high $\tau_{160}$ residuals correspond to H$\alpha$ emission
on large scales, the detailed distribution is not a particularly good
match. The rank correlation coefficient between H$\alpha$ and \htwofir\ over
the area observed by SEST is $\sim 0.1$, much lower than the $0.7$ relating
\htwofir\ and CO. H$\alpha$ emission is proportional to $\int n^2 dl$ and so
obviously a flawed tracer of the true warm gas column ($\int n dl$), but the
poor correspondence on small scales still argues that most \htwofir\ is not
actually warm ionized gas.

{\em Dust Processing in Molecular Clouds:} A significant but hard-to-constrain
uncertainty in Equation \ref{NH2EQ} is if and how dust properties vary between
N83 and the surrounding ISM. The most likely variations are increases in the
FIR emissivity or the $DGR$. In the Milky Way, the FIR emissivity of dust
($\tau_{\rm FIR} / A_V$) does appear to increase towards dense regions,
increasing by $\sim 30$--$50\%$ above $A_V \sim 1$~mag
\citep[e.g.,][]{ARCE99,DUTRA03,CAMBRESY05}. \citet{CAMBRESY01},
\citet{STEPNIK03}, and \citet{CAMBRESY05} argue that this is due to the
creation of fluffy dust grains with low albedos \citep{DWEK97} via grain-grain
coagulation or accretion of gas. At the same time, build-up of existing grains
in molecular clouds and dust creation in Type II supernovae or stellar winds
\citep[e.g.,][]{DWEK98} may cause the $DGR$ ratio near star-forming regions to
be higher than in the surrounding ISM.

The magnitude of grain growth in GMCs remains very poorly constrained and in
an active environment like N83 it will be balanced against grain destruction
(e.g., in shocks). Further, the high dust temperatures, low integrated
extinctions ($A_V \lesssim 2$~mag almost everywhere), and weak CO emission in
N83 are a far cry from the high density, high extinction environments in which
grain coagulation or the formation of icy mantles are usually modeled or
observed \citep[e.g.,][]{OSSENKOPF94}. Moreover, as pointed out by
\citet{BERNARD08}, increased emissivity in Milky Way clouds is often
associated with diminished small grain emission \citep[e.g.,][]{SCHNEE08},
while N83 exhibits increased $I_{70}/I_{160}$ compared to its surroundings.

Because it is unclear what, if any, grain processing is at work in N83, we
make no correction to the emissivity. If dust in N83 indeed has a high
emissivity compared to the diffuse ISM, we will derive values of both
$N(\htwofir )$ and $A_V$ that are too high. Note that our adopted $DGR$ is
already twice that in the surrounding diffuse gas. Increasing or decreasing
the adopted $DGR$ will not affect $\tau_{160}$ or $A_V$, but will lower or
raise $N(\htwofir )$.

{\em Effect of Changing $DGR$ on the CO-H$_2$ relation:} The exact value of
the $DGR$ in N83 is the largest systematic uncertainty in our analysis. We
discuss the constraints on this quantity in \S \ref{N83DGR}. In the right
panel of Figure \ref{WARMGASANDDGR}, we illustrate the relationship between
H$_2$ and $DGR$ in the limiting cases: $DGR$ equal to that in the diffuse ISM
of the SMC Wing (black) and $DGR$ equal to three times this value (gray).

There are two main conclusions to draw from this comparison. First, the
existence and magnitude of a truly CO-free H$_2$ envelope (the $y$-intercept
of the points) depends sensitively on the adopted $DGR$; the lowest plausible
value is (partially by construction) consistent with no envelope and the
highest value implies an envelope with surface densities $\sim
200$--$400$~M$_\odot$~pc$^{-2}$, $1$--$2$ times the average surface density of
a Galactic GMC. Second, the average CO-to-H$_2$ conversion factor varies
between $10$ and $100$ times Galactic over the full range of possible
$DGR$. The qualitative behavior (meaning the presence of bright $I_{\rm CO}$
only above a certain $\Sigma_{\rm H2}$ threshold) remains the same. We
emphasize that the relationship between $A_V$ (or $\tau_{160}$) and $I_{\rm
  CO}$ is unaffected by the choice of $DGR$.

\end{document}